\documentclass[manuscript]{acmart}

\usepackage{graphicx}
\usepackage{balance}  
\usepackage[noend]{algorithmic}
\usepackage[ruled,lined,boxed,commentsnumbered]{algorithm2e}
\usepackage{subfigure}
\usepackage{mdframed}
\usepackage{float}
\usepackage{subfloat}
\usepackage{tikz}
\usetikzlibrary{shapes,arrows,fit,shapes.misc}
\usepackage{verbatim}
\usepackage{multirow}
\usepackage{color}
\usepackage{comment}
\usepackage{graphicx}
\usepackage{enumitem}
\usepackage{hyphenat}
\usepackage{xcolor}
\usepackage{mathtools}
\AtBeginDocument{%
  \mathchardef\mathcomma\mathcode`\,
  \mathcode`\,="8000 
}
{\catcode`,=\active
  \gdef,{\mathcomma\discretionary{}{}{}}
}

\allowdisplaybreaks
\newcommand{\ignore}[1]{{}}

\newcommand{\mE}{\mathcal{E}}

\newcommand{\mI}{\mathcal{I}}
\newcommand{\mL}{\mathcal{L}}

\newcommand{\mR}{\mathcal{R}}

\newcommand{\mV}{\mathcal{V}}
\newcommand{\bA}{\mathbf{A}}

\newcommand{\bDiag}{\mathbf{diag}}
\newcommand{\bM}{\mathbf{M}}
\newcommand{\bX}{\mathbf{X}}
\newcommand{\bY}{\mathbf{Y}}
\newcommand{\andor}{and${\allowbreak}/{\allowbreak}$or}

\newcommand{\rpathsim}{RelSim} 
\newcommand{\relexp}{RRE}
\newcommand{\bmid}{\,\mid\,}

\newcommand{\walkto}[1]{\allowbreak\rightsquigarrow_{#1}\allowbreak}

\newcommand{\bto}{\allowbreak\to\allowbreak}
\newcommand{\bcdot}{\allowbreak\cdot\allowbreak}

\newcommand{\bdashtt}{\allowbreak\texttt{-}\allowbreak}
\newcommand{\bequal}{\allowbreak = \allowbreak}
\newcommand{\bin}{\allowbreak\in\allowbreak}
\newcommand{\bslash}{{\allowbreak}/{\allowbreak}}
\newcommand{\bwedge}{\allowbreak\wedge\allowbreak}
\newcommand{\bvee}{\allowbreak\vee\allowbreak}

\newcommand{\inst}[1]{\mathtt{Inst}(#1)} 
\newcommand{\instv}[2]{\mathtt{Inst}_{#2}(#1)} 
\newcommand{\Sl}[1]{#1} 

\newcommand{\bangle}[1]{\allowbreak\langle #1 \rangle\allowbreak}
\newcommand{\bskip}[1]{\lceil\lceil #1 \rfloor\rfloor}
\newcommand{\btravel}[3]{#2 \hookrightarrow_{#1} #3}
\newtheorem{theorem}{Theorem}
\newtheorem{definition}{Definition}
\newtheorem{corollary}{Corollary}

\newtheorem{proposition}{Proposition}
\newtheorem{example}{Example}

\setcopyright{acmcopyright}
\copyrightyear{2018}
\acmYear{2018}
\acmDOI{10.1145/1122445.1122456}

\acmConference[Woodstock '18]{Woodstock '18: ACM Symposium on Neural Gaze Detection}{June 03--05, 2018}{Woodstock, NY}
\acmBooktitle{Woodstock '18: ACM Symposium on Neural Gaze Detection, June 03--05, 2018, Woodstock, NY}
\acmPrice{15.00}
\acmISBN{978-1-4503-XXXX-X/18/06}

\begin{document}

\title{Structural Generalizability: The Case of Similarity Search}

\author{Yodsawalai Chodpathumwan}
\authornote{Works done while at the University of Illinois at Urbana-Champaign}
\email{yodsawalai.c@tggs.kmutnb.ac.th}
\affiliation{%
\institution{King Mongkut's University of Technology North Bangkok}
\city{Bangkok}
\country{Thailand}
}

\author{Arash Termehchy}
\email{termehca@oregonstate.edu}
\affiliation{%
\institution{Oregon State University}
\city{Corvallis}
\state{OR}
\country{USA}
}

\author{Stephen A. Ramsey}
\email{stephen.Ramsey@oregonstate.edu}
\affiliation{%
\institution{Oregon State University}
\city{Corvallis}
\state{OR}
\country{USA}
}

\author{Aayam Shresta}
\email{shrestaa@oregonstate.edu}
\affiliation{%
\institution{Oregon State University}
\city{Corvallis}
\state{OR}
\country{USA}
}

\author{Amy Glen}
\email{glena@oregonstate.edu}
\affiliation{%
\institution{Oregon State University}
\city{Corvallis}
\state{OR}
\country{USA}
}

\author{Zheng Liu}
\email{liuzhen@oregonstate.edu}
\affiliation{%
\institution{Oregon State University}
\city{Corvallis}
\state{OR}
\country{USA}
}

\begin{abstract}
Graph similarity search algorithms usually leverage the structural properties of a database. 
Hence, these algorithms are effective only on some structural variations of the data and are ineffective on other forms, 
which makes them hard to use.
Ideally, one would like to design a data analytics algorithm 
that is structurally robust, 
i.e., it returns essentially the same accurate results 
over all possible structural variations of a dataset.
We propose a novel approach to create a structurally robust similarity search algorithm over graph databases.
We leverage the classic insight in the database literature 
that schematic variations are caused by having constraints in the database. 
We then present {\rpathsim} algorithm which is provably structurally robust under these variations.
Our empirical studies show that our proposed algorithms are 
structurally robust while being efficient and as effective as or more effective than the state\hyp{}of\hyp{}the\hyp{}art similarity search algorithms.
\end{abstract}

\maketitle


\section{Introduction}
\label{sec:introduction}
\begin{figure*}[t]
\centering
\subfigure[Fragments of DBLP]{
\includegraphics[width=0.405\textwidth]{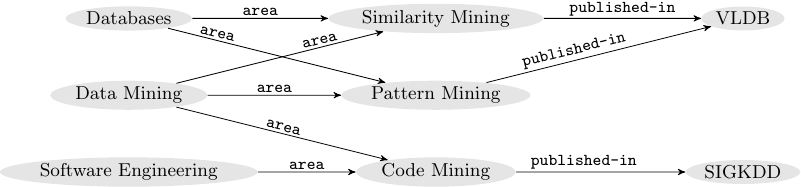} 
\label{fig:mvd-example1-bib}}
\subfigure[Fragments of SIGMOD Record]{
\centering
\includegraphics[width=0.405\textwidth]{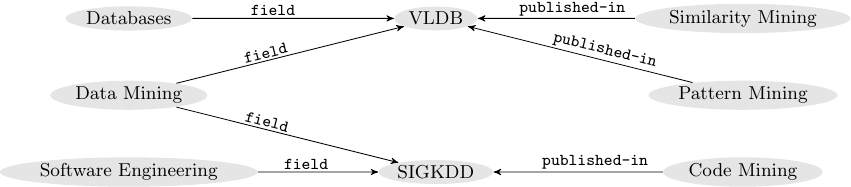}
\label{fig:mvd-example2-bib}}
\vspace{-12pt}
\caption{Example of two bibliography databases whose nodes are papers, conferences and research areas.
}
\label{fig:mvd-example-bib}
\vspace{-14pt}
\end{figure*}

Unsupervised and supervised machine learning (ML) methods are widely used over structured data  \cite{Jeh:KDD:02,Tong:KDD:06,Tong:ICDM:06,Vishwanathan::JMLR:10,Sun:VLDB:11,Zhang:2015:PFT:2783258.2783267,Tang:KDD:2017,DBLP:journals/pvldb/SunZHWCAL20}.
To deliver effective predictions, these methods usually leverage structural or topological 
features of the data that quantify relationships between entities.  
For example, consider the DBLP dataset ({\it dblp.uni-trier.de}) that stores information about papers, conferences, and research areas in computer sciences 
whose fragments are shown in Figure~\ref{fig:mvd-example1-bib}. 
Assume that a user wants to find the most similar research area to {\em Data Mining} in this dataset according to their conferences and publications.
Similarity search algorithms usually use the structure of the graph to measure the degree of similarity between entities in the data.
For example, 
SimRank \cite{Jeh:KDD:02} is a well-known unsupervised similarity search algorithm over graphs that 
quantifies the similarity between two entities according to how likely two random surfers will meet each other if they start from the two entities \cite{Jeh:KDD:02,Tong:KDD:06,Tong:ICDM:06,Vishwanathan::JMLR:10,Sun:VLDB:11,Zhang:2015:PFT:2783258.2783267}.
Thus, it correctly finds {\em Data Mining} to be more similar to {\em Databases} than to {\em Software Engineering} in Figure~\ref{fig:mvd-example1-bib}.
Oversimplifying a bit, since {\em Data Mining} and {\em Databases} share relatively more neighbors, i.e., papers, over this dataset, they have a relatively high SimRank score.
These similarity measures, such as SimRank, are used as features to build ML models for various data analytics tasks, e.g., pattern query matching or community detection \cite{Antonellis:VLDB:08,Tong:KDD:06,Tong:ICDM:06,Zhang:2015:PFT:2783258.2783267}.

It is, however, established that different datasets usually represent essentially the same information in different forms and structures \cite{AliceBook,infopreserve:hull,Fagin:TCS:05,DBLP:conf/sigmod/MelnikAB07,infopreserve:XML,NORM,XMLFD,Barcelo:ICDT:13,Francis:2017:SMD:3034786.3056113,Boneva:2015:EDBTGraphQ}.
A classic and well-known example of such structural variations is (de-)normalization in relational and XML data where the original and normalized databases represent the same information under different 
structures \cite{AliceBook,NORM,XMLFD,Arenas:2006:NTX:1228268.1228284}. 
As another example, consider the SIGMOD Record bibliographic dataset
({\it sigmod.org$\bslash$\\publications}) whose fragments are shown in Figure~\ref{fig:mvd-example2-bib}. 
Intuitively, this dataset represent essentially the same information about the same set of {\it paper}, 
{\it conference}, and {\it research area} entities as the one in Figure~\ref{fig:mvd-example1-bib}. 
But, each dataset has its own way of representing these entities and their relationships. 
For example, DBLP connects directly each paper to its research areas and conferences. 
Given that all papers in a conference share the same set of research areas, 
one can also choose the structure in Figure~\ref{fig:mvd-example2-bib} to represent this information and connect research areas of a paper to its conferences.
This dataset represents the relationship between the research area of a paper using a path through the conference of the paper instead of a direct link as in Figure~\ref{fig:mvd-example1-bib}.
We have been working with bioinformatics experts to analyze large graph datasets from various data sources and 
have observed that they often restructure their graph data to meet certain efficiency-oriented or data quality goals. 
For instance, if two nodes are far apart in the data and are often queried together in some lookup queries, 
the experts add new edges that connect these entities directly to rewrite and run lookup queries faster. 
This modification does {\it not} add any new information to the database as the new edge can be inferred from the current 
(long) path between the corresponding entities.

Structural features delivers different values across the aforementioned variations, which may return to inaccurate results on some structures for the same information.
For example, SimRank finds {\em Data Mining} more similar to {\em Software Engineering} than {\em Databases} on the data fragment in Figure~\ref{fig:mvd-example2-bib}, 
which is clearly not accurate. 
As explained in the preceding paragraph, Figure~\ref{fig:mvd-example1-bib} and \ref{fig:mvd-example2-bib} contain essentially the same information about the relationships between the same set of entities.
Thus, the accuracy of SimRank depends on the structure and the form information is presented 
rather than the content of the underlying data.
Due to the constant data evolution \cite{pagerank:stability,Ghoshal:NatureComm:2011,DBLP:journals/cacm/StonebrakerFDB17,integration:evolving,oodb:evolving} 
and growing need for analyzing different datasets with a great deal of structural variations \cite{DBLP:conf/sigmod/BalazinskaCAFKS20,DBLP:conf/vldb/2018polydmah,DBLP:journals/debu/StonebrakerI18,Francis:2017:SMD:3034786.3056113,DBLP:conf/sigmod/MelnikAB07,Arenas:2009:RSM:1620585.1620589,Boneva:2015:EDBTGraphQ,DBLP:conf/sigmod/FernandezDMQTAE17}, 
similarity features and models are often used across datasets with different forms and structures. 
Thus, the accuracy of current similarity features and models is unstable and unpredictable, e.g.,  
they may perform well on the datasets used to develop or train these features but poorly on others.

It is known that variations in the content of the data across different datasets may reduce 
the efficacy of features and models when applied on a new dataset, e.g., due to their bias toward certain data distributions 
\cite{Mitchell97,10.5555/645531.655828,10.1145/1772690.1772790,Shamir82clusterstability,Lange:NIPS:03,David:COLT:06}. 
Thus, it has been a central problem in ML and data analysis 
to measure or improve the robustness or generalizability of features and models 
against variations in the content of datasets \cite{Mitchell97,10.5555/645531.655828,Rigollet:JMLR:07,Shamir82clusterstability}.
For the same reason, it is also important to ensure that similarity models and algorithms generalize and 
are effective to {\it unseen structures}, i.e., 
the structure of the information other than the ones used to develop or train features and models.
For example, consider a developer who is building a system that finds similar nodes in the input graph datasets. 
To develop and choose similarity features and algorithms, she may test feature on some datasets and select the one
that returns the answers that the domain experts deem accurate. Users would like to use this system on datasets other than the ones
used to test its effectiveness whose may follow a different style of structure than the datasets used to test the system.
It may not be, however, possible for the developer to test the system over sufficiently many datasets that cover all potential
structural variations that users may face when using the system.
Thus, it is not clear whether the developed features and algorithms for similarity will be effective to the datasets over
which the system is used by users. This will significantly limit the off-the-shelf use of such systems.

The robustness and generalizability of features
in the face of structural variations across different datasets have not been explored. 
In this paper, we investigate the robustness of similarity features and algorithms in the face of the structural variations of the data. 
We focus on the problem of similarity search as it is both a popular type of queries over graphs and an important building block of other data analysis tasks, such as pattern query matching, community detection, and clustering \cite{Tong:KDD:06,Tong:ICDM:06,Jeh:KDD:02,Zhang:2015:PFT:2783258.2783267}.
We have two main goals in our approach to finding robust similarity features and algorithms.
First, we aim at building a theoretical framework that captures a sufficiently general notion of structural variations
and robustness, which is able capture a wide variety of structural variations. 
We leverage the rich body of research on structure and schema management and variation in database community 
to achieve this goal. Second, instead of developing new features and algorithms that can generalize and are robust
against structural variations, we would like to extend and enhance current features and algorithms to be robust. 
Since current methods and features are currently used in data analytics systems, 
this approach will make it easier to achieve robustness in current systems. 
Moreover, this effort may lay the foundation and provide insights on how to enhance current features or algorithms used in 
other ML and data analysis tasks to achieve structural robustness.

In our investigations in this paper, 
we establish a relationship between the structural robustness of similarity search features and algorithms 
and the languages used to specify these features in the graph, e.g., the language that expresses the 
types of paths used for random walks in the graph. 
That is, oversimplifying a bit, the more expressive the language is, the more robust feature similarity algorithm is.
Interestingly, this is in contrast with the robustness and generalizability of models against statistical variations in the content of the data where usually a more expressive model may be harder to generalize to unseen data than a less expressive one \cite{Mitchell97}.
It is very time-consuming to compute values for features that use an overly expressive language \cite{Gutierrez:JCSS:11,Wood:2012:QLG:2206869.2206879}.
Thus, we propose a feature specification language that is sufficiently expressive to be robust over 
popular structural variations and its expressed features are computed efficiently over large data.
Moreover, as it may be hard for users to work with a complex language to express features, 
we propose a usable method in which users specify high-level hints for the system to find and compute relevant features.
In particular, we make the following contributions.
\begin{itemize}[noitemsep,nolistsep,leftmargin=*]
\item 
We define structural generalizability and robustness of a similarity search algorithm. (Section~\ref{sec:repindep}).
To find and explore structural variations of a graph database, we leverage results in database literature, which state that database constraints give rise to its structural variations and extend it to graph data \cite{4568391,DBLP:journals/tods/FaginInverse,infopreserve:hull}.

\item
We show that existing similarity search features and 
algorithms are {\it not} robust because of the limited types of relationships they use to measure the degree of similarity. 
We propose a robust algorithm called {\it RelSim}, 
which extends a well-known similarity search algorithm 
({\it PathSim} \cite{Sun:VLDB:11,shi2014hetesim}) by using a sufficiently expressive set of patterns.
We also show that features expressed in such language are computed efficiently over large data.

\item
RelSim requires users to specify the relationship between nodes to compute similarity score between the nodes using the proposed language, which may be relatively hard. We propose an algorithm that leverages simple guidelines from the user to compute structurally robust similarity scores efficiently (Section~\ref{sec:robustalg-simple}).

\item 
We report the results of our extensive empirical studies 
over large graph databases, which indicate that our proposed algorithms are structurally robust and improves the effectiveness of its original similarity search algorithm 
(Section~\ref{sec:experiment}). 
Our empirical studies also show that our algorithms are efficient over large data.
\end{itemize}


\ignore{
For example, it is not clear whether a similarity feature or algorithm, such as SimRank, 
generalizes and performs effectively across datasets with various 

We extend this result for graph databases. For example, the variation in Figure~\ref{fig:mvd-example-bib} will {\it not} preserve the information in the database in Figure~\ref{fig:mvd-example1-bib} unless all papers in a conference share the same set of research areas. 
Otherwise, it is {\it not} clear what research areas must connect to a conference in Figure~\ref{fig:mvd-example2-bib}.
We show that given some mild assumptions, the constraints that induce structural variations over graph databases are in the form of tuple-generating dependencies (tgd), which are a well-known family of constrains over databases \cite{AliceBook,Boneva:2015:EDBTGraphQ}.

We show that the aforementioned variations modify the patterns and features used by current similarity search algorithms to compute the degree of similarity between nodes in the graph database. We also show that current similarity search algorithms will be robust in the face of these structural variations if they use a set of relatively more complex patterns to compute similarity scores. This is in fact an interesting result as so far the main justification for using a rich set of relationships and patterns in a graph analysis algorithm has been to improve the accuracy of its result.
This result indicates that such an approach may make the algorithm structurally robust, too. 
More specifically, we extend a well-known similarity search algorithm, called PathSim \cite{Sun:VLDB:11,shi2014hetesim}, using a set of sufficiently complex pattern to compute robust similarity score efficiently.
}
\ignore{
Hence, to use a data analytics algorithm, the user has to restructure the database to find the structure
over which the algorithm delivers accurate results. 
Since there is {\it not} any clear guideline on how to find such a desirable structure for the algorithm, one has to do this through trial and error, which takes a great deal of time and effort.

There are robust algorithms over certain types 
of schematic variations 
\cite{Picado:2017:SIR:3035918.3035923,Tang:KDD:2017,Truong2012,ChodpathumwanAT16}. 
They, however, have two major shortcomings. 
First, they are robust only over a subset of 
frequently occurring schematic variations. 
Because they leverage the properties special to the variation 
over which they are robust, it is {\it not} clear how to generalize these algorithms 
to be robust against other schematic variations. 
Second, current schematically robust systems generally 
either propose new algorithms~\cite{Truong2012}, 
or make significant and/or complex modifications to the current ones  
\cite{Picado:2017:SIR:3035918.3035923}.     
However, current algorithms have 
been widely adapted and it is costly to replace them with new ones.
Hence, one should aim at making current algorithms robust to schematic variations using simple modifications.
Moreover, current algorithms are shown to be effective over some data representations. 
Thus, a robust version of them will be effective over more representations.

One generic method is to run an algorithm and over all possible variations of a validation subset of the database and select 
the representation with the most accurate answers. 
Nonetheless, databases have a large number of possible structural variations 
\cite{infopreserve:XML,AliceBook}.
For example, a relational table may have exponential number of normalizations.  
Moreover, such a validation subset is not generally available 
for unsupervised methods, such as similarity search. 
This approach also requires the underlying database be transformed 
to the desired representation, which may not be practical for a large 
{\andor} constantly evolving databases. 
}

\section{Graph Data and Constraints}
\label{sec:datamodel}
We fix a countably infinite set of node ids denoted by $\mV$.
Let $\mL$ be a finite set of labels. 
A {\em database} $D$ over $\mL$ is a directed graph $(V, E)$ 
in which $V$ is a finite subset of $\mV$ and 
$E \subseteq V \times L \times V$.
This definition of graph databases is frequently used in the graph data management literature
\cite{Barcelo:ICDT:13,Wood:2012:QLG:2206869.2206879,Fan:2016:FDG:2882903.2915232}.
We denote an edge from node $u$ to node $v$ 
whose label is $a$ as $(u, a, v)$.
We say that $(u, a, v) \bin D$ whenever $(u,a,v) \in E$.
Similarly, we say that $v \in D$ whenever $v \in V$.

{\em Database constraints} restrict the instances of a schema. 
They are usually expressed as logical formulas 
over schemas 
\cite{Boneva:2015:EDBTGraphQ,Beeri:1984:PPD:1634.1636,AliceBook}.
Two widely used type of constraints are
{\em tuple-generating} and {\em equality-generating} dependencies  \cite{Boneva:2015:EDBTGraphQ,Beeri:1984:PPD:1634.1636}.
A tuple-generating dependency ({\it tgd} for short) over schema $\mL$ 
is in the form of 
$\forall \bar{x} (\phi(\bar{x}) \bto \exists \bar{y} \psi(\bar{x}, \bar{y}))$ 
where $\bar{x}$ and $\bar{y}$ are sets of variables, 
and $\phi$ and $\psi$ are logical formulas in a query language over $\mL$.  
A {\it full tgd} does {\it not} have any existential variable in its conclusion.
An equality\hyp{}generating dependency ({\it egd} for short) over schema $\mL$ is in the form of $\forall \bar{x} (\phi(\bar{x}) \bto x_1 = x_2)$ where $\bar{x}$ is a set of variables, $x_1, x_2$ $\in \bar{x}$, and $\phi$ is logical formulas in a query language over $\mL$.
\begin{example}
\label{ex:constraint}
Database shown in 
Figure~\ref{fig:mvd-example1-bib} contains a constraint 
$(x_1,\texttt{area},x_3) 
\bwedge (x_3,\texttt{pub}\bdashtt\texttt{in},x_4)$ 
$\bwedge (x_2,\texttt{pub}\bdashtt\texttt{in},x_4) 
\bto (x_1,\texttt{area},x_2)$.
\end{example}

Tgds and egds are arguably the most popular and frequently used types 
of database constraints and generalize popular constraints, such as function and multi\hyp{}valued dependencies \cite{AliceBook}. 

A commonly studied query language over graph databases is 
{\em conjunctive regular path queries} (conjunctive RPQ), 
which is used to express tgd and egd constraints over graph databases 
\cite{Francis:2017:SMD:3034786.3056113,Cruz:87:SIGMOD,Wood:2012:QLG:2206869.2206879,Barcelo:ICDT:13}.
The {\em RPQ} $p$ over schema $\mL$ is defined by the following grammar:
\[
p := \; 
\epsilon \bmid a\; (a \bin \mL) \bmid a^-\; (a \bin \mL) 
\bmid p \cdot p \bmid p + p \bmid p^*
\]
\noindent
in which $\epsilon$ is an empty label, 
$^-$ is a reverse traversal of an edge, 
$\cdot$ is a concatenation, 
$+$ is a disjunction, and 
$*$ is a Kleene star.
To avoid parentheses and ambiguity, 
it is assumed that the reverse traversal has the highest priority,
then Kleene star, then concatenation and then disjunction. 
Example of an RPQ over a schema of a database shown in 
Figure~\ref{fig:mvd-example2-bib}
is $\texttt{field}\bcdot\texttt{published}\bdashtt\texttt{in}^-$.
The RPQ $p$ defines a binary relation over database nodes.
More precisely, the result of evaluating $p$ on database $D$ 
is a set of pairs of nodes in $D$ such that 
there is a path defined by $p$ between the two nodes. 
We denote the result of evaluating $p$ over $D$ as $[[p]]_D$.
For example, given label $a$ in the schema of $D$, 
the result of $[[a]]_D$ is a set of pairs of nodes $\{(u,v)\}$ 
where there is an edge with label $a$ from $u$ to $v$. 
Let $\bar{x}=$ $(x_1,\ldots,x_n)$ and $\bar{y}=$ $(y_1,\ldots,y_m)$ 
be tuples of distinct variables. 
A conjunctive RPQ is a formula $\phi(\bar{x})$ of the form 
$\exists \bar{y} ((z_1, p_1, z'_1) \land ... \land (z_k, p_k, z'_k))$ 
where $p_i$ is an RPQ and 
$z_i, z'_i \in$ $\{x_1, \ldots, x_n,y_1,\ldots,y_m\}$, $1 \leq i \leq k$ 
\cite{Wood:2012:QLG:2206869.2206879,Barcelo:ICDT:13}.
We call $(z_i, p_i, z'_i)$ an {\em atom} of $\phi(\bar{x})$.

A schema $S$ is a pair $(\mL,\Gamma_S)$ in which 
$\mL$ is a (finite) set of labels and 
$\Gamma_{S}$ is a finite set of constraints.
By the abuse of notation, 
we say that a label $l \bin S$ if $l \bin \mL$ and
a constraint $\gamma \bin S$ if $\gamma \bin \Gamma_S$.
Each {\it database} of schema $S \bequal (\mL, \Sigma_S)$ is a database over $\mL$ such that all constraints in $\Sigma_{S}$ holds.
We denote the set of all databases of schema $S$ as $\inst{S}$.
A similarity query $q$ over database $I(V_I, E_I) \in \inst{S}$ is a node id in $V_I$ \cite{Sun:VLDB:11,WeirenYu:SIGIR:12,Tong:ICDM:08,Jeh:KDD:02,Zhao:CIKM:09,Jeh:WWW:03,He2010PSC,Antonellis:VLDB:08,shi2014hetesim}.
The answers to a similarity query $q$ over $I$ is a ranked list of node ids in $I$ that are not equal to $q$.

\section{Structural Robustness and Variations}
\label{sec:repindep}

\subsection{Structural Robustness}
\label{sec:robustness}

Intuitively, a structurally\hyp{}robust query answering algorithm 
should return essentially the same (list of) answers for the same query 
across databases that contain the same information content. 
Researchers have leveraged the concept of invertible transformation 
to formalize the equivalence of information stored in different databases 
\cite{infopreserve:hull,infopreserve:XML}. 
A {\it transformation} from schema $S$ to schema $T$ 
is a function from $\inst{S}$ to $\inst{T}$, 
which maps each database of $S$ to a database of $T$ \cite{infopreserve:hull,infopreserve:XML}. 
We denote a transformation from $S$ to $T$ as $\Sigma_{ST}$. 
For example, a transformation $\Sigma_{1a,1b}$ from schema of the database in Figure~\ref{fig:mvd-example1-bib}
to the schema of the one in Figure~\ref{fig:mvd-example2-bib}  changes the structure of the database in Figure~\ref{fig:mvd-example1-bib} such that the research areas associated with a paper become connected to the paper via the conference of the paper 
and produces the database in Figure~\ref{fig:mvd-example2-bib}.

This definition of transformations may {\it not} be sufficiently powerful to capture structural variations. Assume that the schema of the database in Figure~\ref{fig:mvd-example2-bib} contains an additional type of relationship called {\it keyword-of} that connects each paper published in a conference to nodes, which store keywords of the paper. Consider an updated version of the database in Figure~\ref{fig:mvd-example2-bib} in which each paper is connected to some additional nodes via relationship {\it keyword-of}. Intuitively, the updated database has more information than the one in Figure~\ref{fig:mvd-example1-bib}. Let us define transformation $\Sigma_{1a,1c}$ between the schema in Figure ~\ref{fig:mvd-example2-bib} and the updated schema such that it modifies relationships between research areas, conferences, and papers similar to $\Sigma_{1a,1b}$ and adds some keywords to each paper.
This transformation maps each database in the original schema to multiple databases under the transformed schema where each database may have different (number of) keywords for the same paper.
Thus, we use a definition of transformation between schemas $S$ and $T$ in which the transformation $\Sigma_{ST}$ establishes a relation between 
$\inst{S}$ and $\inst{T}$ to cover the aforementioned cases \cite{Fagin:TCS:05,Barcelo:ICDT:13}. It maps each database $I \in \inst{S}$ to at least one database $ J \in\inst{T}$ and it is {\it not} a function. One denotes the fact that $J$ is a transformation $I$ under $\Sigma_{ST}$ as $(I,J) \models \Sigma_{ST}$ \cite{Fagin:TCS:05,Barcelo:ICDT:13}. For brevity and by the abuse of notation, we show the transformed databases of $I$ under $\Sigma_{ST}$ as
$\Sigma_{ST}(I)$.

The transformation $\Sigma_{ST}$ is {\em invertible} if there is a transformation $\Sigma_{TS}$ from $T$ to $S$ such that, for each database $I \in \inst{S}$, $\Sigma_{TS}$ maps every database $\Sigma_{ST}(I)$ to $I$ and only $I$, i.e., for each database in $\Sigma_{ST} (I)$, $\Sigma_{TS}( \Sigma_{ST} (I))$ is exactly $I$. 
In other words, the composition of $\Sigma_{ST}(I)$ and $\Sigma_{TS}(I)$, shown as $\Sigma_{ST}(I) \circ$ $\Sigma_{TS}$ is equivalent to 
the identity transformation $id$ that maps each database only to itself.
In this case, we call $\Sigma_{TS}$ the {\it inverse} of $\Sigma_{TS}$ and denote it as $\Sigma_{ST}^{-1}$. 
If there is an invertible transformation from schema $S$ to $T$, one can reconstruct exactly the information in each database $I$ from the information available in $\Sigma_{ST}(I)$. In other words, each database in $\Sigma_{ST}(I)$ has at least the same amount of information as $I$. 
%
We say that $S$ has at least as much information as $T$ and denote their relationship as $S \stackrel{\Sigma_{ST}}{\preceq} T$ or simply $S \preceq T$ if $\Sigma_{ST}$ is clear from the context.

\begin{example}\label{ex:invertibleSet}
Consider transformation $\Sigma_{1a,1b}$ from schema of the database in Figure~\ref{fig:mvd-example1-bib}
to the schema of one in Figure~\ref{fig:mvd-example2-bib}. 
Because of the constraint described in Example~\ref{ex:constraint}, this transformation is invertible.
Intuitively, this constraint over Figure~\ref{fig:mvd-example1-bib} implies that 
papers published in the same conference are related to the same set of research areas.
Hence, one may change the structure shown in Figure~\ref{fig:mvd-example1-bib} such that the research areas associated with a paper is instead connected to the paper via the conference of the paper and get the database fragment shown in Figure~\ref{fig:mvd-example2-bib}.
One can recover the information in the original database using the information in Figure~\ref{fig:mvd-example2-bib}. One can find the exact set of research areas directly connected to each paper in Figure~\ref{fig:mvd-example1-bib} by checking the research areas directly connected to the conference of that paper.
Similarly, $\Sigma_{1a,1c}$ is also invertible as one can follow the same approach to recover the information of the database in Figure~\ref{fig:mvd-example1-bib} from the transformed databases.
%
\end{example}

Our definition of inverse extends the notion of Fagin-inverse used in the context of relational data exchange \cite{DBLP:journals/tods/FaginInverse}. 
In a Fagin-inverse, a transformation may map multiple databases from schema $S$ to multiples ones under schema $T$ and its inverse maps multiple databases under $T$ back to multiple ones under $S$.
The answers to a similarity query in different databases of $S$ may be different.
Hence, we use a more strict version of inverse because 
our goal is to compare the results of an algorithm over a single database 
and all its structural variations that contain the same information.
Therefore, we are interested in the transformations that map a single database under the original schema to one or more databases under the transformed schema. 
Consequently, the inverse transformation must map those databases under the transformed schema to the original database and only the original database. 
Therefore, for the rest of this paper, we assume that each transformation maps a single database to multiple ones and its inverse maps multiple databases back only to the original database.   

Next, we present the definition of a {\em structurally robust} (robust for short) algorithm.
Roughly speaking, a robust algorithm must return the same results for the same input query over a database and databases under invertible transformations.
Two (ranked) lists of node ids are {\em equivalent} if they contain exactly the same node ids at the same positions. 
Two empty lists of answers are equivalent.
\begin{definition}\label{def:robustness}
Given schemas $S$ and $T$ such that $S \stackrel{\Sigma_{ST}}{\preceq} T$, an algorithm is {\it robust} under $\Sigma_{ST}$ 
if it returns equivalent answers for every input query $q$ over every database $I \in \inst{S}$ and every database in $\Sigma_{ST}(I)$.
\end{definition}
\noindent
An algorithm is robust under a set of transformations if it is robust under all members of the set. 

\subsection{Structural Variations}
\label{sec:variation}

\subsubsection{Expressing Transformations}
\label{sec:variation-epressing}

To characterize the structural variations of a schema, one needs to express invertible transformations. 
Researchers usually use {\em declarative (schema) mappings} to express schematic variations in graph and relational databases \cite{Fagin:TCS:05,DBLP:journals/tods/FaginInverse,Barcelo:ICDT:13}.
Roughly speaking, a transformation between schemas $S$ and $T$ is expressed as a set of logical formulas $\phi_{S}(\bar{x})$ $\bto \psi_{T}(\bar{y})$ where $\phi_S(\bar{x})$ and $\psi_T(\bar{y})$ are queries over schemas $S$ and $T$, respectively. 
More precisely, transformation $\Sigma_{ST}$ between graph schemas $S$ and $T$ is a finite set of rules $\phi_S(\bar{x}) \bto \psi_T(\bar{y})$ such that $\bar{x} \subseteq$ $\bar{y}$ and $\phi_{S}(\bar{x})$, i.e., {\em premise}, and $\psi_T(\bar{y})$, i.e., {\em conclusion}, are conjunctive RPQs over $S$ and $T$, respectively \cite{Barcelo:ICDT:13}. 
In each rule $\phi_S(\bar{x}) \bto \psi_T(\bar{y})$, every variable in $\bar{x}$ is universally quantified and every one in the set of $\bar{y}$ either belongs to $\bar{x}$ or is existentially quantified.

\begin{example}\label{ex:infopreservingT}
\vspace{-3pt}
The transformation $\Sigma_{1a,1b}$ from schema of the database in Figure~\ref{fig:mvd-example1-bib}
to the one in Figure~\ref{fig:mvd-example2-bib} is expressed using a mapping with two rules:  
$(x_1,\texttt{pub-in},x_2) \bto (x_1,\texttt{pub-in},x_2)$ and 
$(x_1,\texttt{area}\cdot\texttt{pub-in},x_2) \bto (x_1,\texttt{field},x_2)$. 
The inverse of $\Sigma_{1a,1b}$ can also be expressed using following rules:
$(x_1,\texttt{field}\cdot\texttt{pub}\bdashtt\texttt{in}^-,x_2) 
\bto (x_1,\texttt{area},x_2)$ 
and 
$(x_1,\texttt{pub}\bdashtt\texttt{in},x_2) 
\bto (x_1,\texttt{pub}\bdashtt\texttt{in},x_2)$.
The transformation $\Sigma_{1a,1c}$ that maps databases under the schema in Figure~\ref{fig:mvd-example1-bib} to the one shown in Figure~\ref{fig:mvd-example2-bib} with some keyword nodes connected to each paper published in a conference can be expressed using the aforementioned rules plus rule 
$(x_1,\texttt{pub-in},x_2) \bto (y_1,\texttt{keyword-of},x_1)$, 
where $y_1$ is an existential variable that represents the nodes that contain keywords of the paper $x_1$. This rule can map a database to multiple one each of which has different number of keyword nodes.
The inverse of $\Sigma_{1a,1c}$ is the same as the inverse of $\Sigma_{1a,1b}$ as it does {\it not} need the keywords to reconstructs the data.
\vspace{-3pt}
\end{example}

Transformation $\Sigma_{ST}$ maps each database $I \bin \inst{S}$ to $J\bin \inst{T}$ if for each rule $\phi_{S}(\bar{x}) \bto \psi_{T}(\bar{y})$ in $\Sigma_{ST}$, we have $\bar{u} \bin [[\psi_{T}(\bar{x})]]_J$ if $\bar{u} \in [[\phi_{S}(\bar{y})]]_I$ \cite{Barcelo:ICDT:13}.
We use the closed world semantic for schema mappings \cite{Hernich:2011:CWD:1966385.1966392}. That is, transformation $\Sigma_{ST}$ maps $I \bin \inst{S}$ to only databases whose nodes and edges are constructed using the mapping rules.
The alternative semantic is the open world semantic in which the transformations of $I \bin \inst{S}$ under $\Sigma_{ST}$ may contain nodes and edges that are {\it not} created as the result of the schema mapping rules \cite{Fagin:TCS:05,DBLP:journals/tods/FaginInverse,Barcelo:ICDT:13}.
A consequence of this semantic is that the inverse of transformation $\Sigma_{ST}$ will map each database $\Sigma_{ST}(I)$ to other databases in addition to $I$, e.g., all databases in $S$ that include at least the same amount of information as $I$ \cite{DBLP:journals/tods/FaginInverse}. 
Our goal, however, is to compare the results of similarity queries over the original database $I$ and its variations. Thus, we use the closed world semantic for schema mappings. 

\subsubsection{Information Preservation}
\label{sec:variation:characterize}

Next, we present the full characterization of invertible transformations of a schema. It helps us to identify the set of structural variations of a schema, which we use to design robust algorithms. 
Consider transformation $\Sigma_{ST}$ from schemas $S$ to $T$ and $\Sigma_{TR}$ from schemas $T$ to $R$. 
The {\em composition} of $\Sigma_{ST}$ and $\Sigma_{TR}$, denoted as $\Sigma_{TR} \circ$ $\Sigma_{ST}$, is a transformation from $S$ to $R$ such that 
if $J \in \Sigma_{ST}(I)$ and $K \in \Sigma_{TR}(J)$, then $K \in (\Sigma_{TR} \circ \Sigma_{ST})(I)$. 

Given transformations $\Sigma_{ST}$ from schema $S$ to $T$, its inverse $\Sigma_{ST}^{-1}$ is a transformation from $T$ back to $S$.
Thus, the composition $\Sigma_{ST}^{-1} \circ \Sigma_{ST}$ will map the database $I\in \inst{S}$ to (only) itself.
In other words, the composition of $\Sigma_{ST} \circ$ $\Sigma_{ST}^{-1}$ are a set of rules, i.e., constraints, over $S$.
We have the following proposition by applying the definitions of the inverse and composition of transformations.

\begin{proposition}\label{prop:composition}
Given schemas $S$ and $T$ such that $S \stackrel{\Sigma_{ST}}{\preceq} T$, for all databases $I \in \inst{S}$  
we have $I \models \Sigma_{ST}^{-1} \circ \Sigma_{ST}$.
\end{proposition}
\noindent
Proposition~\ref{prop:composition} extends the results on the lossless decomposition of a relational schema \cite{4568391} 
and the ones on the inverse of a relational schema mapping in \cite{DBLP:journals/tods/FaginInverse}.
According to Proposition~\ref{prop:composition}, 
the constraints on schema $S$ determine its structural variations. 
Following this proposition, the constraints on $S$ that give rise to invertible structural variations are in tgds. 

\begin{example}\label{ex:SourceConstraints}
The composition $\Sigma_{1a,1b} \circ \Sigma_{1a,1b}^{-1}$ in Example~\ref{ex:infopreservingT} results in a constraint 
$(x_1,\texttt{area},x_4) \bwedge (x_4,\texttt{published}\bdashtt\texttt{in}, x_3) 
\bwedge (x_2,\texttt{published}\bdashtt\texttt{in},x_3)
\bto (x_1, \texttt{area}, x_2)$ which is equivalent to the constraint of the database shown in Figure~\ref{fig:mvd-example1-bib} 
as described in Example~\ref{ex:constraint}.
\end{example}

We should note that the composition of two transformations may {\it not} be expressible using first order schema mapping formulas \cite{Fagin:2005:CSM:1114244.1114249,Barcelo:ICDT:13}.
Roughly speaking, each rule in $\Sigma_{TR} \circ$ $\Sigma_{ST}$ is created by replacing an atom $(x,exp,y)$ in the premise of a rule in $\Sigma_{TR}$ by the premise of a rule in $\Sigma_{ST}$ whose conclusion matches $(x,exp,y)$ \cite{Fagin:2005:CSM:1114244.1114249,Barcelo:ICDT:13,Boneva:2015:EDBTGraphQ}. 
Assume that the conclusion of a rule in $\Sigma_{ST}$ contains an atom $(x,exp,y)$ with an existentially quantified variable, which may be stated explicitly in the rule or introduced by using concatenation, Kleene star, or disjunction, i.e., $+$, in $exp$. 
Also, assume that there is an atom in the premise of $\Sigma_{ST}^{-1}$ that matches $(x,exp,y)$. 
In this case, one needs second-order logic to express this composition \cite{Fagin:2005:CSM:1114244.1114249}. If this happens to a transformation and its inverse, then the set of constraints on the original schema will be tgds in second order logic.
To the best of our knowledge, there has {\it not} been any work on database constraints over graph (or relational) databases that are in languages more expressive than the first order logic, e.g., second order logic, and the database constraints in the first order logic are by far more widely used than the ones expressed in higher order logics \cite{Boneva:2015:EDBTGraphQ,AliceBook}. 
Thus, 
we focus our attention to the first order constraints as explained in Section~\ref{sec:datamodel}.
If the atoms in the premise of $\Sigma_{ST}^{-1}$ match the atoms with only universally quantified variables in the conclusion of $\Sigma_{ST}$, their compositions can be expressed using the (first order) tgds as defined in Section~\ref{sec:datamodel} \cite{Fagin:2005:CSM:1114244.1114249,Barcelo:ICDT:13}. 
Therefore, we consider only transformations whose composition with their inverses can be expressed in first order logic.

As shown in Example~\ref{ex:SourceConstraints}, the composition of transformation $\Sigma_{1a,1b}$ and its inverse and transformation $\Sigma_{1a,1c}$ and its inverse can be expressed as such tgds. 
In this case the shared atoms between the premises of rules in $\Sigma_{ST}^{-1}$ and the conclusions of rules in  $\Sigma_{ST}$ will be in form of $(x, l, y)$ or $(x, l^{-1}, y)$ where $l$ is a label in schema $T$.
Thus, each rule in $\Sigma_{ST}^{-1} \circ$ $\Sigma_{ST}$ is created by replacing an atom $(x,l,y)$ in the premise of a rule in $\Sigma_{ST}^{-1}$ by the premise of each rule in $\Sigma_{ST}$ whose conclusion matches $(x,l,y)$ (or $(x, l^{-1}, y)$ by exchanging the positions of $x$ and $y$). This method naturally extends to the atoms of the form $(x, l^{-1}, y)$ in the premise of rules in $\Sigma_{ST}^{-1}$.

Consider a transformation $\Sigma_{ST}$ from $S$ to $T$ where the conclusions of a rule $\alpha$ in $\Sigma_{ST}$ contain an existentially quantified variable $z$. Given $I \in \inst{S}$, the nodes in databases $\Sigma_{ST}(I)$ created as the result of applying $\alpha$ to $I$ and correspond to $z$ do {\it not} have any fixed ID (or value if applicable) as they do {\it not} correspond to any node in $I$. 
Thus, $\Sigma_{ST}(I)$ will contain more than a single database. Since a transformation maps a database to multiple ones, its inverse must map multiple databases to a single one. 
Thus, the inverse can be expressed using a set of rules without any existentially quantified variable in their conclusions.
Similar results have been shown for the structure of similar types of inverse of schema mappings over relational databases \cite{DBLP:journals/tods/FaginInverse}.
Thus, the composition of $\Sigma_{ST}$ and $\Sigma_{ST}^{-1}$ is a set of rules where the premise of each rule is a conjunctive RPQ and its conclusion is a single RPQ atom in form of $(x, exp, y)$ where $exp$ is either $l$ or $l^{-}$ where $l$ is a label in $S$. Hence, $\Sigma_{ST}^{-1} \circ$  $\Sigma_{ST}$ is a set of full tgds over $S$. 

The set of tgds introduced by Proposition~\ref{prop:composition} is necessary to have invertible transformations for schema $S$, but it is {\it not} sufficient. We show that $S$ must satisfy an additional group of tgds to have invertible variations. 
Let $\sigma$ denote the set of tgd constraints in $\Sigma_{ST}^{-1} \circ \Sigma_{ST}$.
Given $\sigma$, we create another group of tgd constraints over $S$, denoted as $\sigma^*$, as follows.
For each tgd constraint in $\sigma$ whose conclusion is in the form of $\chi_1(x,y) \bto (x,l^{-}, y)$, we replace it with constrain $\chi_1(y,x) \bto (y, l, x)$. 
Then, for all tgds with the same atom in their conclusions, i.e., $\chi_1(x,y) \bto (x, l, y), ..., \chi_m(x,y) \bto (x, l, y)$ in $\sigma$, 
we construct the constraint $(x, l, y) \bto \chi_1(x,y) \bvee ... \bvee \chi_m(x,y)$. 
For each label $l'$ in $\mL_S$ that does {\it not} appear in a conclusion 
of any constraint in $\sigma$, we create the constraint 
$(x, l', y) \bto \mathtt{FALSE}$, which means that there is {\it not} any database in $\inst{S}$ with any edge whose label is $l'$.

\begin{proposition}\label{theorem:SourceConstraints}
Given transformations $\Sigma_{ST}$ from $S$ to $T$ 
and $\Sigma_{TS}$ from $T$ to $S$, let $\sigma$ denote $\Sigma_{ST} \circ \Sigma_{TS}$. 
$\Sigma_{ST}$ is invertible with inverse $\Sigma_{TS}$ if and only if, for every database $I \in \inst{S}$, 
we have $I \models \sigma \land \sigma^*$.
\end{proposition}
In the rest of this paper, we refer to an invertible transformation simply as a transformation.

\ignore{
Table ~\ref{framework:notation} summarizes the notations used in our framework.

\begin{table}
    \centering
    \caption{Summary of the notations.}
    \vspace{-10pt}
    \label{framework:notation}
    \begin{tabular}{|l|p{6.0cm}|}
    \hline
    Notation & Definition\\
    \hline
    $\inst{S}$ & Databases of schema $S$\\
    \hline
    $\Sigma_{ST}$ & Transformation from schema $S$ to $T$ \\
    \hline
    $\Sigma_{ST}(I)$ & Transformations of $I$ based on $\Sigma_{ST}$\\
    \hline
    $\Sigma_{ST}^{-1}$ & The inverse of $\Sigma_{ST}$ \\
    \hline
    $\Sigma_{ST} \circ \Sigma_{TR}$ & The composition of transformations\\
    \hline
    \end{tabular}
\end{table}
}

\subsubsection{Computing Inverses}
\label{sec:variation:inverse}
There have been numerous works on computing the inverses of a schema mapping over relational databases
\cite{DBLP:journals/tods/FaginInverse,DBLP:journals/tods/FaginKPT08,Arenas:2009:RSM:1620585.1620589}.
In this paper, we focus on analyzing the robustness of algorithms over transformations and 
designing algorithms that are robust against them. The theoretical results presented in the preceding 
section enable us to characterize these transformations and design robust algorithms over them.
We do not need to compute the inverse of a transformation in this paper. Thus,
the problem of computing an inverse of an input schema mapping is not the subject of this paper and is an interesting future work. To provide the reader with some insight on the problem of computing an inverse of 
a graph schema mapping, we extend the following result from the relational schema mappings. 
The proof uses the results of and is similar to the one of Theorem 14.9 in \cite{DBLP:journals/tods/FaginInverse}.
\begin{theorem}
Given transformation $\Sigma_{ST}$ from $S$ to $T$, 
the decision problem of whether 
$\Sigma_{ST}$ is invertible is coNP-hard.
\end{theorem}

\section{Robust Similarity Search}
\label{sec:robustalg}

\ignore{
\begin{figure}[t]
\centering
\subfigure[Blibiographic DB]{
\includegraphics[width=0.5\textwidth]{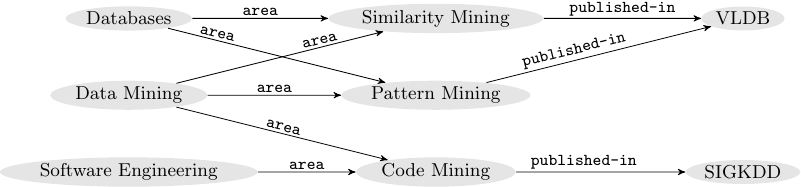} 
\label{fig:mvd-example1-bib}}

\subfigure[Alternative representation of Blibiographic DB]{
\centering
\includegraphics[width=0.5\textwidth]{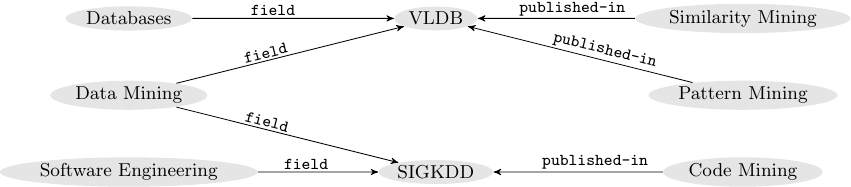}
\label{fig:mvd-example2-bib}}
\caption{Example of two bibliography databases.
}
\label{fig:mvd-example-bib}
\end{figure}
}

\subsection{Robustness of Current Methods}

To the best of our knowledge, frequently used 
structural\hyp{}based similarity search features and algorithms are 
based on random walks, 
e.g., RWR \cite{Tong:ICDM:06}, 
pairwise random walk, e.g., 
SimRank \cite{Jeh:KDD:02} and P-Rank \cite{Zhao:CIKM:09}, or 
path-constrained framework, e.g., 
PathSim \cite{Sun:VLDB:11} and HeteSim \cite{shi2014hetesim}.
There are also other similarity search algorithms that extend the aforementioned algorithms 
such as common neighbors, $Katz_{\beta}$ measure, commute time, and sampled set of random paths/walks between nodes
\cite{katz1953new,Zhang:2015:PFT:2783258.2783267}.
The algorithms that are not path-constrained, e.g., SimRank and RWR, 
are mainly developed to measure similarity over graphs where
all edges have the same label \cite{Jeh:KDD:02,Tong:ICDM:06}.
However, these algorithm are sometimes used over data graphs with multiple edge labels \cite{Sun:VLDB:11}.
They have also been used as a basis for methods developed for data graphs with multiple labels \cite{Sun:VLDB:11,shi2014hetesim}.
We consider both the original and the extended versions of SimRank and RWR, for the sake of completeness.

However, some people have modified and used these two algorithms over databases with multiple types of entities and relationship types \cite{Sun:VLDB:11}.
In this paper, we consider the extended version of SimRank and RWR.

Similarity scores computed by algorithms that use random walks 
and pairwise random walks are largely influenced by the 
topology of the graph. Because some transformations may modify
the topological structure of a database, 
structural\hyp{}based similarity search algorithms 
such as RWR and SimRank are not robust under these
variations as shown in our empirical studies
in Section~\ref{sec:experiment}.
Similar to these algorithms, there are algorithms that leverage the idea of random walks
by randomly picking some walks or paths between two nodes and randomly traversing certain number steps on each walk \cite{Zhang:2015:PFT:2783258.2783267}.
The more similar two nodes are, the more likely it is for one to reach to one of them by starting from another one using the aforementioned method.
As they use similar core ideas to RWR and SimRank to measure the similarity between two nodes, their results are similarly influenced by transformations that modify the database topology. 
An invertible transformation may change the length and the number of paths and walks between two nodes. 
For example, the length of the path between nodes {\it VLDB} and {\it Pattern Mining} is one in Figure~\ref{fig:mvd-example1-bib} and two in its variation 
(Figure~\ref{fig:mvd-example2-bib}). 
One may significantly reduce or increase the length of paths between two entities in a database and its variations under similar transformation. 
Thus, these methods may deliver different similarity scores for the same pairs of nodes over a 
a database and the database under an invertible transformation.

Two entities may be similar based on the paths or patterns that separate them in a database where those paths or patterns represent a certain type of relationship.
The degree of similarity between two entities may largely depend on the type of relationship between them. 
For instance, consider a database with researchers, conferences in which they publish, and their affiliations. 
Some users may want to find similar researchers from the point of view of their affiliations. Other users may like to find similar researchers based on the conferences that they publish their papers in.
Thus, one often has to consider the type of relationship between two entities to define an effective similarity metric with a clear semantic.
Path-constrained similarity search algorithms, such as PathSim and HeteSim, follow this approach \cite{Sun:VLDB:11,shi2014hetesim}. 
They allow users to supply a path template, or {\em pattern}, that specifies the type of relationship between entities in their queries.
Example of a pattern in Figure~\ref{fig:mvd-example-bib} is 
$m_1:$ $\texttt{pub-in}$ $\bcdot$ $\texttt{pub-in}^-$, which
reflects the relationship between two papers through a conference in which they are both published in.
Each instance of a pattern in database $D$ is a path in $D$ whose sequence of 
edge labels match the sequence of labels in the pattern. 
For example, {\it Similarity Mining}$\bcdot$$\texttt{pub-in}$$\bcdot${\it VLDB}$\bcdot$$\texttt{pub-in}^{-}$$\bcdot${\it Pattern Mining} is an instance of $m_1$
in Figure~\ref{fig:mvd-example2-bib}.
The PathSim score of entities $u$ and $v$ in database $D$ given a pattern $p$ is  
\begin{equation}\label{eq:pathsim}
\texttt{sim}_p(u,v,D) \bequal \frac{2 \times |u \walkto{p} v|}{|u \walkto{p} u| + |v \walkto{p} v|}
\end{equation}
where $|u \walkto{p} v|$, $|u \walkto{p} u|$ and $|v \walkto{p} v|$ 
denote the numbers of $(u,p,v)$, $(u,p,u)$ and $(v,p,v)$ path instances in $D$, respectively.
The robustness of PathSim or other path-constrained similarity search methods largely depend on the representation of the underlying relationships.

\begin{example}\label{ex:why}
Consider two representations of bibliographic data in Figure~\ref{fig:mvd-example-bib}.
Suppose a user wants to find similar research areas to {\em Data Mining} based on their shared conferences.
In Figure~\ref{fig:mvd-example1-bib}, the user uses the pattern $p_1:$ 
\texttt{area}$\bcdot$
\texttt{published}\hyp{}\texttt{in} $\bcdot$
\texttt{published}\hyp{}\texttt{in}$^-$ $\bcdot$
\texttt{area}$^-$
to represent the relationship and compute similarity scores between research areas.
PathSim then finds {\em Data Mining} more similar
to {\em Databases} than to {\em Software Engineering}.
However, in Figure~\ref{fig:mvd-example2-bib}, the same user may use the pattern $p_2:$ 
\texttt{field}$\bcdot$
\texttt{field}$^-$
to compute similarity scores between research areas. 
This pattern, however, finds that both {\em Software Engineering} and {\em Databases} are equally similar to {\em Data Mining}.
\end{example}

\subsection{Achieving Robustness}
\label{sec:sr-pathsim}

Example~\ref{ex:why} illustrates that 
there may {\it not} be any (obvious) pattern over some structure or schema of a dataset 
to express the desired relationship between entities.
If a user wants to find the similarity of two research areas based on their shared conferences, 
she can use pattern $p_2$ over the structural representation in Figure~\ref{fig:mvd-example2-bib}, 
but she cannot find the same pattern in Figure~\ref{fig:mvd-example1-bib}.
One may propose a candidate pattern $p_1$ over Figure~\ref{fig:mvd-example2-bib}.
However, it includes more information than $p_2$, i.e., 
information about the set of papers published in the conferences can be captured by $p_1$, but not by $p_2$.

On the other hand, the user may like to measure the similarity of research areas 
based on their shared conferences and also the papers published in those conferences.
Over Figure~\ref{fig:mvd-example1-bib}, she can use pattern $p_1$ to help measure the similarity.
However, there is no pattern that expresses that relationship in Figure~\ref{fig:mvd-example2-bib}.

One may solve this problem by using a language that is more expressive than the sequence of relationship labels to express relationship types between entities in a database.

\begin{example}\label{ex:idea-skip}
\vspace{-1pt}
Following Example~\ref{ex:why}, one can create an equivalent relationship
to the one expressed by $p_2$ in Figure~\ref{fig:mvd-example1-bib} 
by modifying $p_1$ to treat the set of all paths through some papers 
from a conference to a research area as a single path, 
i.e, {\it skip} details of entities visited along those paths.
\end{example}

The resulting pattern from Example~\ref{ex:idea-skip} 
considers only the existence of a connection between a research area and a conference in the database 
as opposed to $p_1$ that takes into account all papers that connect a research area to a conference. 
This pattern intuitively represents an equivalent relationship over
Figure~\ref{fig:mvd-example1-bib} to the one conveyed by $p_2$ over Figure~\ref{fig:mvd-example2-bib}. 
Hence, one has to define and add an operation that implements the aforementioned skipping behavior to the language 
that describes relationships between entities. 

\begin{example}\label{ex:idea-nested}
\vspace{-1pt}
Following Example~\ref{ex:why}, 
one can modify pattern $p_2$ such that the a that follows the pattern 
can visit the publications of conference while visiting a conference. 
This way, we will get a relationship between two research areas 
that takes into consideration both the conferences and publications shared between them 
in Figure~\ref{fig:mvd-example2-bib}. 
The resulting pattern 
expresses an equivalent relationship to what $p_1$ represents over Figure~\ref{fig:mvd-example1-bib}. 
\end{example}
However, even following this approach, 
one should be careful not to increase the expressivity of the relationship language too much 
as it takes a long time to find all instances of a complex pattern and compute its similarity score in a large database.

We present a relationship expressing language that is expressive enough to represent 
equivalent relationships across various representations of the same datasets. 
We also show that using this relationship language, 
there is a similarity algorithm that returns equal similarity scores between every pair of corresponding entities over different representations of the same information.
More precisely, an algorithm that computes a similarity score 
using Equation~\ref{eq:pathsim} where 
$p$ is written in our proposed language is structurally robust.

To implement the solution presented in Example~\ref{ex:idea-nested}, 
one may use the idea of nested operation in the nested regular expression (NRE) language \cite{Barcelo:ICDT:13}. 
Let $[p]$ denote a nested path of $p$ where 
a path $(u,[p],u)$ exists if and only if there exists a node $v$
such that a path $(u,p,v)$ exists.
To achieve the same results as $p_1$ over Figure~\ref{fig:mvd-example1-bib}, 
the user should use the pattern $p_4 :$ 
\texttt{field}$\bcdot$
$[\texttt{published-in}^-]$ $\bcdot$
$[\texttt{published-in}^-]$ $\bcdot$
\texttt{field}$^-$
to compute a similarity score between research areas.
That is, similar research areas are based on shared conferences, 
and the strength of this relationship is based on
the number of publications published in those conferences.

Next, We define the extension to NRE 
namely {\em rich\hyp{}relationship expression} ({\relexp}), 
over  schema $S$ as 
\[ 
p := 
\epsilon \bmid a\; (a \bin \Sl{S}) \bmid p^- \bmid p^* 
\bmid p \cdot p \bmid p + p \bmid [p] \bmid \bskip{p}
\]
\noindent
where $[\;]$ denotes a nested operation and $\bskip{\;}$ denotes a skip operation.

Since Equation~\ref{eq:pathsim} used the number of instances 
of a specified pattern when calculating the similarity score, 
we define an {\em instance} of an {\relexp} as follows.
An instance of some {\relexp} in a graph database $D$ 
is a ternary relation $(u,v,s)$ representing a traversal over $D$
from node $u$ to node $v$ whose actual traversal are recorded in a sequence $s$.
Each entry in the recorded sequence $s$ is either 
a node id, an edge label or a string of pattern.
Equivalence between two {\relexp} instances is defined 
naturally by entry-wise comparison.

Given a sequence $s = \bangle{s_1,...,s_m}$ and 
$t = \bangle{t_1,...,t_n}$ of $m$ and $n$ entries, respectively, 
let $s \bullet t = \bangle{s_1,...,s_m,t_2,...,t_n}$
which is defined only if $s_m = t_1$;
and let $\bar{s} = \bangle{\grave{s_m},...,\grave{s_1}}$
where, for each $i = 1...m$, 
$\grave{s_i} = s_i$ if $s_i$ represents a node
and $\grave{s_i} = s_i^-$ otherwise.
A set of instances of an {\relexp} $p$
in a database $D$ in schema $S$, denoted by $\mI_D(p)$, is 
defined as follows.
For a given label $a \bin \Sl{S}$, 
arbitrary {\relexp}s $p$, $p_1$ and $p_2$
over $S$, we have 
\begin{align*}
\mI_D(\epsilon) = \{ & (u,u,\bangle{u}) \bmid u \text{ is a node in } D \} \\
\mI_D(a) 		= \{ & (u,v,\bangle{u,a,v}) \bmid (u,a,v) \bin D \} \\
\mI_D(p^-)		= \{ & (v,u,\bar{s}) \bmid (u,v,s) \bin \mI_D(p) \} \\
\mI_D(p_1 \cdot p_2)    = \{ 	& (u,v,s_1 \bullet s_2) \bmid \forall w, (u,w,s_1) \bin \mI_D(p_1) \\
						        &  \text{ and } (w,v,s_2) \bin \mI_D(p_2) \} \\
\mI_D(p_1 + p_2)        = \{ 	&(u,v,s) \bmid (u,v,s) \bin \mI_D(p_1) \cup \mI_D(p_2) \} \\
\mI_D(p^*)      = \{ & (u,v,s) \bmid (u,v,s) \bin \mI_D(\epsilon) \cup \mI_D(p) \cup \mI_D(p^2) \cup ... \} \\
\mI_D(\bskip{p}) = \{	& (u,v,\bangle{u,\widetilde{p},v}) \bmid \exists s, (u,v,s) \bin \mI_D(p) \} \\
\mI_D([p]) 		= \{ 	& (u,u,s \bullet \bangle{v,u}) \bmid \forall v, (u,v,s) \bin \mI_D(p) \}
\end{align*}
\noindent 
where 
$\widetilde{p}$ is a string  $p$ with all $\bskip{\;}$ removed, 
e.g., $\widetilde{\bskip{a \cdot b}} = a \cdot b$.
We define the definition of instances of 
an {\relexp} for a particular pair of nodes $u$ and $v$ 
in database $D$ such that
\[
\mI^{u,v}_D(p) \bequal \{ (u,v,s) \bmid \forall s, (u,v,s) \bin \mI_D(p) \}.
\]
\noindent
If database $D$ is clear from the context, 
we may write $\mI^{u,v}_D(p)$ and $\mI_D(p)$ simply as 
$\mI^{u,v}(p)$ and $\mI(p)$, respectively.
For the remaining of this paper, we assume all relationship patterns are {\relexp}s. 
\begin{proposition}\label{prop:pathexp}
Given a schema $S$, 
$a \bin S$, $p$, $p_1$ and $p_2$ 
are arbitrary {\relexp}s over $S$,
and a database $D \bin \inst{S}$
where nodes $u$ and $v$ are in $D$, 
the following properties hold.
\begin{enumerate}[label=(\arabic*),nolistsep,noitemsep] 
\item \label{prop:item:1}
If $\mI^{u,v}_D(p) \neq \emptyset$,
then $|\mI^{u,v}_D(\bskip{p})| = 1$. \\
Otherwise, $|\mI^{u,v}_D(\bskip{p})| = 0$. 
\item \label{prop:item:2}
$\mI^{u,v}_D(\bskip{a}) \bequal \mI^{u,v}_D(a)$
\item \label{prop:item:3}
$|\mI^{u,v}_D(p_1 \cdot p_2)| \bequal
\sum_{w \bin D} {|\mI^{u,w}_D(p_1)||\mI^{w,v}_D(p_2)|}$  
\item \label{prop:item:4}
If $(u,p_1,v) \bin D$ iff $(u,p_2,v) \bin D$,
then $|\mI^{u,v}_D(\bskip{p_1})| \bequal
|\mI^{u,v}_D(\bskip{p_2})|$. 
\item \label{prop:item:5}
$|\mI^{u,u}_D([p])| \bequal |\mI^{u,u}_D(p \cdot \bskip{p^-})|$
\end{enumerate}
\end{proposition}

Given a transformation
$\gamma: \phi(\bar{x}) \bto (x_1,a,x_2)$, 
one can construct an undirected graph 
$G_\gamma = (V,E)$ such that 
$V = \{\bar{x}\}$ and 
$E$ is a set of an edge $(x_i,p,x_j)$ where 
$(x_i,p,x_j)$ is an atom in $\phi(\bar{x})$.
We say that $\gamma$ is acyclic if $G_\gamma$
contains {\it no} cycle. 
In the following theorem, we assume that the premise of every transformations are acyclic.
Using Proposition~\ref{prop:pathexp}, we prove the following theorem
by induction on 
the number of disjunctions and concatenated labels 
in a given {\relexp} pattern.

\begin{theorem}\label{thrm:existpathexp}
Given schemas $S$ and $T$, for every invertible transformation $\Sigma_{ST}$ and every pattern $p$ over $S$, 
there exists a pattern $p'$ over $T$ such that, for every database $D \bin \inst{S}$ and $J \bin \Sigma_{ST}(D)$
$\forall u,v \bin D$, $|\mI^{u,v}_{D}(p)| \bequal |\mI^{u,v}_{J}(p')|$.
\end{theorem}


\begin{proof}
For~\ref{prop:item:1} and~\ref{prop:item:2}, 
the statements hold directly from definitions of path instances.
For~\ref{prop:item:3}, proofs are done by counting.
For~\ref{prop:item:4}, assume $\exists (u,p_1,v) \in D$.
We have $(u,p_1,v) \bin D$ iff $(u,p_2,v) \bin D$, 
and so $\mI^{u,v}_D(p_1) \neq \emptyset$ iff $\mI^{u,v}_D(p_2) \neq \emptyset$.
That is, $|\mI^{u,v}_D(\bskip{p_1})| \bequal 1$ 
iff $|\mI^{u,v}_D(\bskip{p_2})| \bequal 1$.
Otherwise, $\mI^{u,v}_D(p_1) \bequal \mI^{u,v}_D(p_2) = \emptyset$,
and so $|\mI^{u,v}_D(\bskip{p_1})| \bequal |\mI^{u,v}_D(\bskip{p_2})| \bequal 0$.
For~\ref{prop:item:5}, by definitions, 
$(u,p,v) \bin D$ iff $(u,\widetilde{p},v)$ iff $(v,\widetilde{p^-},u)$.
Hence, $|\mI^{u,u}_D([p])| 
\bequal
|\{ (u,u,s \bullet \bangle{v,u}) \bmid \forall v, (u,v,s) \bin \mI_D(p) \}|
\bequal
|\{ (u,u,s \bullet \bangle{v,\widetilde{p^-},u}) \bmid \forall v, (u,v,s) \bin \mI_D(p) \}|
\bequal 
|\{ (u,u,s \bullet \bangle{v,\widetilde{p^-},u}) \bmid \forall v, (u,v,s) \bin \mI_D(p) 
\text{ and } (v,u,\bangle{v,\widetilde{p^-},u}) \bin \mI_D(\bskip{p^-}) \}|
\bequal |\mI^{u,u}_D(p \cdot \bskip{p^-})|$.
\end{proof}


Then, we prove the theorem.

\begin{proof}
Assume each node in a graph database has node ids; 
and those same ids always exist in any solution under any invertible schema mapping.


If every label in the pattern $p$ exists in both 
schemas $S$ and $T$, we have that, 
for each label $a \in S$ appearing in $p$, 
$\forall u',v' \bin D$, 
$(u',a,v') \bin D$ iff $(u',a,v') \bin \Sigma_{ST}(D)$.
Clearly, 
$|\mI^{u,v}_D(p)| \bequal 
|\mI^{u,v}_{\Sigma_{ST}(D)}(p)|$.

Suppose $p = \bskip{r}$ for some pattern $r$ over $S$.
By Proposition~\ref{prop:pathexp}\ref{prop:item:4}, 
if there exists a pattern $r'$ over $T$ s.t.
$|\mI^{u,v}_D(r)|>0$ iff $|\mI^{u,v}_{\Sigma_{ST}(D)}(r')|>0$,
we have 
$|\mI^{u,v}_D(p)| \bequal |\mI^{u,v}_{\Sigma_{ST}(D)}(\bskip{r'})|$.
Also, by Proposition~\ref{prop:pathexp}\ref{prop:item:5},
one may write $p \cdot \bskip{p^-}$ instead of $[p]$.
Hence, we may consider a pattern $p$ 
without any use of $\bskip{\;}$ or $[\;]$.
Further, since $\mI(p^*) = \mI(\epsilon) \cup \mI(p) \cup \mI(p^2) \cup ...$, 
if there exists $p'$ such that 
$|\mI^{u,v}_D(p)| = |\mI^{u,v}_{\Sigma_{ST}(D)}(p')|$,
then 
$|\mI^{u,v}_D(p^*)| = |\mI^{u,v}_{\Sigma_{ST}(D)}(p'^*)|$.
Hence, we may also consider a pattern $p$ without the use of $*$.

Assume $p = p_1+...+p_m$ 
where $p_1,...,p_m$ are distinct and contain no `$+$'.
That is $\mI_D(p_i) \cap \mI_D(p_j) = \emptyset$ for any $i \neq j$.
Clearly, $|\mI_D(p)| = |\mI_D(p_1)|+...+|\mI_D(p_m)|$.

We first show that, for each $i=1...m$,  
there exists a pattern $p'_i$ over $T$ s.t. 
$\forall u,v \bin D$, 
$|\mI^{u,v}_D(p_i)| \bequal |\mI^{u,v}_{\Sigma_{ST}(D)}(p'_i)|$
using strong induction over the number of concatenations in $p_i$.

Clearly, if $p_i = a$ or $p_i = a^-$ 
where $a \bin S$ and $a \bin T$, 
then the statement holds.
Otherwise, since $\Sigma_{ST}$ is information preserving, 
there exists a transformation rule in its inverse s.t.   
$\phi(x_1,x_2,\bar{x}) \bto (x_1,a,x_2)$.
One can then construct a pattern $p'_i$ 
that traverses $\phi(\bar{x})$ from $x_1$ to $x_2$.
We have that, $\forall u,v \bin D$, 
$(u,p_i,v) = (u,a,v) \bin D$ iff $\phi(u,v,\bar{x})$
iff $(u,p'_i,v) \bin \Sigma_{ST}(D)$.
Let $p''_i = \bskip{p'_i}$.
By Proposition~\ref{prop:pathexp}\ref{prop:item:4}, 
$|\mI^{u,v}_D(p_i)| = |\mI^{u,v}_{\Sigma_{ST}(D)}(p''_i)|$.
The proof extends for the case where $p = a^-$. 
\ignore{
$\phi_1(x_1,x_2,\bar{x}) \bto (x_1,a,x_2)$, ...,
$\phi_k(x_1,x_2,\bar{x}) \bto (x_1,a,x_2)$.
Because each rule is acyclic, one can construct
a pattern $p'_{i,j}$ 
that traverses $\phi_j(\bar{x})$ from $x_1$ to $x_2$
for each $j=1...k$.
We have that, $\forall u,v \bin D$, 
$(u,a,v) \bin D$ iff $\bigvee_{j=1...k} \phi_j(u,v,\bar{x})$
iff $(u,p'_{i1}+...+p'_{ik},v) \bin \Sigma_{ST}(D)$.
Let $p'_i = \bskip{p'_{i1}+...+p'_{ik}}$.
By Proposition~\ref{prop:pathexp}\ref{prop:item:4}, 
$|\mI^{u,v}_D(p_i)| = |\mI^{u,v}_{\Sigma_{ST}(D)}(p'_i)|$.
The proof extends for the case where $p = a^-$. 
}

Suppose the statement holds for any $p_i$
that contains up to $k$ concatenations.
Without losing generality, 
let $p_i = p_{i,1} \cdot p_{i,2}$, 
for some $p_{i,1},p_{i,2} \neq \epsilon$,
containing $k+1$ concatenations.
Hence, $p_{i,1}$ and $p_{i,2}$ contain at most $k$ concatenations.
Consider that, $\forall u,v,w \in D$,
there exists $r_{i1}$ and $r_{i2}$ in $T(D)$ s.t. 
$|\mI^{u,w}_D(p_{i1})| \bequal |\mI^{u,w}_{\Sigma_{ST}(D)}(r_{i1})|$
and 
$|\mI^{w,v}_D(p_{i2})| \bequal |\mI^{w,v}_{\Sigma_{ST}(D)}(r_{i2})|$. \\
Thus $|\mI^{u,v}_D(p)|$ 
$\bequal$ 
$\sum_{w \bin D}\allowbreak{|\mI^{u,w}_D(p_{i1})||\mI^{w,v}_D(p_{i2})|}$ \\
$\bequal$ 
$\sum_{w \bin \Sigma_{ST}(D)}\allowbreak{|\mI^{u,w}_{T(D)}(r_{i1})||\mI^{w,v}_{\Sigma_{ST}(D)}(r_{i2})|}$ 
$\bequal$  
$|\mI^{u,v}_{\Sigma_{ST}(D)}(r_{i1} \cdot r_{i2})|$. \\
That is, $p''_i = r_{i1} \cdot r_{i2}$ satisfies the claim.

\ignore{
Next we show that $p'_j = p_i$, for each $i \neq j$.
Consider if $p'_j = p'_i$, where $i \neq j$, and
there is no other such $p'_j$.
There must exist a transformation rule in the inverse of $\Sigma_{ST}$
that maps to multiple labels in $S$, and there is no rule 
that maps to each of those labels.
Hence, $\Sigma_{ST}$ is not information preserving.
}

Using an induction over the number of disjunction over $p$,
we have that there exists a pattern
$p' = p'_1+...+p'_k$ s.t. the theorem holds.
\end{proof}

\newcommand{\tta}{\texttt{a}}
\newcommand{\ttb}{\texttt{b}}
\newcommand{\ttc}{\texttt{c}}
\newcommand{\ttd}{\texttt{d}}
\newcommand{\tte}{\texttt{e}}
\newcommand{\ttf}{\texttt{f}}
We restrict our attention to transformations with acyclic premises 
in order to reduce the expressivity of the relationship language and keep the computation of similarity scores efficient.
A cyclic premise allows multiple traversals 
from one variable to another in the premise, 
and requires an indicator in the relationship language 
whether two variables along the traversal
are the same, e.g., starting and ending nodes in a cycle are the same.
In this case, it is {\it not} possible to rewrite this pattern over the premise to an equivalent one 
without a conjunction ($\bwedge$).
For instance, consider the pattern representing the premise of a cyclic tgd 
$(x_1, \tta, x_2) \bwedge (x_2, \ttb, x_3) \bwedge (x_3, \ttc, x_4)
\bwedge (x_1, \ttd, x_3) \bwedge (x_2, \tte, x_4) \bto (x_1, \ttf, x_4)$.
Assume we want to rewrite the pattern over this premise similar to $(x_1, exp, x_4)$ for some {\relexp} $exp$.
To remove the conjunction in 
$(x_1, \tta, x_2) \bwedge (x_2, \ttb, x_3)$, 
one may write $(x_1, \tta \bcdot \ttb, x_3)$.
However, because $x_2$ is specified in $(x_2, \tte, x_4)$, 
$x_2$ cannot be removed, and so this conjunction is necessary.
Hence, the language to properly express this relationship pattern
should be a conjunctive {\relexp}.
Since conjunctive {\relexp} is more complex, 
it will take longer to compute the similarity scores between nodes. 
Regardless, the result of Theorem~\ref{thrm:existpathexp} extends for general tgd constraints if a conjunction is added to our proposed relationship expression language.

The following is an immediate result of Theorem~\ref{thrm:existpathexp}.
\begin{corollary}\label{crllry:existpathexp}
Given a database $D$ of a schema $S$, 
for every transformation $\Sigma_{ST}$ for some schema $T$, 
there is a mapping $M$ between 
the set of patterns over $S$ and
the set of patterns over $T$
such that, 
for a given pattern $p$ over $S$, 
we have that 
$\forall D \bin \inst{S}$, $\forall u,v \bin D$, 
$\texttt{sim}_p(u,v,D) \bequal 
\texttt{sim}_{M(p)}(u,v,\Sigma_{ST}(D))$.
\end{corollary}
Corollary~\ref{crllry:existpathexp} guarantees that, 
for each pair of entities $u$ and $v$ and pattern $p$ between them over a dataset, 
one can always find a equivalent pattern with equal similarity score to $p$ 
between $u$ and $v$ on other variations of the database.
Hence, the returned ranked list of answers to a similarity query across databases under this
transformation are always the same. 
We call the algorithm that uses 
Equation~\ref{eq:pathsim} to compute similarity on {\relexp} patterns {\em Relationship-Similarity} 
({\rpathsim}).

One may extend RWR or SimRank so that the similarity measurement  
is based on a particular relationship pattern between entities \cite{Sun:VLDB:11}.
RWR computes a similarity score between nodes $u$ and $v$ 
in a dataset using the steady\hyp{}state probability 
that a random walk from $u$ will stay at $v$.
SimRank, on the other hand, computes the score based on the probability that 
two random walks from $u$ and $v$ are met at a vertex in the data graph.
Technically, the probability of a random walk from $u$ to $v$ 
computes the chance that a walk from $u$ {\em hop}s from a node to its neighbor 
repeatedly until reaching $v$. 
Each hop, hence, is intuitively defined as a single edge between two nodes.
In this extended RWR or SimRank, given a relationship pattern, 
a hop is defined only if a walk 
follows 
the given pattern from one node to another node.
Following this idea, 
we can use the same measurement as SimRank and RWR to compute similarity scores 
over a relationship pattern as similarly specified in {\rpathsim}.
Using a similar proof to Theorem~\ref{thrm:existpathexp}, 
we prove the following proposition.
Let $\texttt{RWR}_p(u,v,D)$ and $\texttt{SimRank}_p(u,v,D)$
denote a similarity score between nodes $u$ and $v$ computed
using RWR and SimRank scoring function that only considers
walks that follows {\relexp} $p$. 
\begin{proposition}\label{prop:rwrsimrank}
Given a database instance $D$ of a schema $S$, 
for every transformation $\Sigma_{ST}$ for some schema $T$, 
there is a mapping $M$ between 
a set of patterns over $S$ and
a set of patterns over $T$
such that, 
for a given pattern $p$ over $S$, 
we have 
$\forall D \bin \inst{S}$, $\forall u,v \bin D$, 
$\texttt{RWR}_p(u,v,D) \bequal 
\texttt{RWR}_{M(p)}(u,v,\Sigma_{ST}(D))$
and 
$\texttt{SimRank}_p(u,v,D) \bequal 
\texttt{SimRank}_{M(p)}(u,v,\Sigma_{ST}(D))$.
\end{proposition}
\begin{proof}
The proof is similar to the one of Theorem~\ref{thrm:existpathexp}.
Specifically, RWR and SimRank assume that 
the weight of connectivity between two nodes in a data graph 
depends on the number of instances of the relationship pattern in the database.
The weight matrices are then used to compute similarity score between two nodes.
\end{proof}

PathSim is shown to be more effective than RWR and SimRank \cite{Sun:VLDB:11}. 
Thus, we focus on our extension of PathSim.

\subsection{Computing Similarity Scores}
\label{sec:langs-computation}

For a pattern with only concatenations, 
the number of {\relexp} instances can be computed using {\em commuting matrix}
\cite{Sun:VLDB:11}.
Given labels $l_1,...,l_m$ in a schema $S$, 
a commuting matrix of pattern $p = l_1 \cdot ... \cdot l_m$ 
over database $D$ is $\bM_p = \bA_{l_1}\bA_{l_2}...\bA_{l_m}$
where $\bA_{l_i}$ is an adjacency matrix that represents
a number of edges of label $l_i$ between pairs of nodes in $D$.
Each entry $\bM_p(u,v)$ represents the number of instances of $p$ 
from node $u$ to node $v$ in $D$. 
Given a commuting matrix, we can compute a similarity score
$\texttt{sim}_p(u,v,D)$ as $\frac{2\bM_p(u,v)}{\bM_p(u,u)+\bM_p(v,v)}$ 
\cite{Sun:VLDB:11}.

We extend the computation of commuting matrix for
{\relexp}s as follows.
Given matrices $\bX$ and $\bY$, 
let $>$ be a boolean operation such that each entry $(i,j)$ 
of $\bX>\bY$ is 1 
if $\bX(i,j)>\bY(i,j)$
or 0 otherwise, and
$\bDiag\{\bX\}$ denote a diagonal matrix of $\bX$. 
Given a label $a$ and arbitrary patterns $p$, $p_1$ and $p_2$ 
over database $D$ in schema $S$, we have
$\bM_{a} \bequal \bA_a$, 
$\bM_{p^-} \bequal \bM_p^T$, 
$\bM_{p_1 \cdot p_2} \bequal \bA_{p_1}\bA_{p_2}$, 
$\bM_{p_1 + p_2} \bequal \bA_{p_1}+\bA_{p_2}$ if $p_1 \neq p_2$, 
$\bM_{p_1 + p_2} \bequal \bA_{p_1} = \bA_{p_2}$ if $p_1 \bequal p_2$, 
$\bM_{\bskip{p}} \bequal \bM_p > \mathbf{0}$, and
$\bM_{[p]} \bequal \bDiag\{\bM_p(\bM_{p}^T>\mathbf{0})\}$
where $\mathbf{0}$ denotes a matrix whose entries are zero.

Since computing a commuting matrix for {\relexp}
$p$ over database $D$ follows standard matrix operations, 
the complexity is bounded by $O(m|V|^3+n|V|^2)$
where $m$ denotes the number of matrix multiplications, e.g.,
the number of concatenations and nested operations in $p$, 
$n$ denotes the number of other operations, 
and $|V|$ denotes the number of nodes in $D$.
Therefore, {\rpathsim} still has the same complexity as that of PathSim.

\section{Simplifying \rpathsim}
\label{sec:robustalg-simple}
The relationship expression language presented in Section~\ref{sec:robustalg} 
may be too complicated for average users.
For instance, $p_4:$ 
\texttt{field}$\bcdot$
$[\texttt{pubslihed-in}^-]\bcdot$
$[\texttt{pubslihed-in}^-]\bcdot$
\texttt{field}$^-$,
which involves nested operations, 
is less intuitive than $p_2:$
\texttt{field}$\bcdot$\texttt{field}$^-$
over Figure~\ref{fig:mvd-example2-bib}, 
which uses only concatenations.
To improve the usability of a similarity search algorithm,
we would like to enable users to submit their patterns using a 
relatively intuitive set of operations, 
such as concatenations and reverse traversals.
Additionally, users may like to measure 
the similarity between entities 
using a set of different types of relationships to get a more holistic and robust 
view of the similarity between entities \cite{Sun:VLDB:11}.
For example, users may want to use both $p_2$ and $p_4$ to compute 
the similarity between research areas in Figure~\ref{fig:mvd-example2-bib}. 

Thus, we propose a robust and effective algorithm
whose input is a pattern that may contain 
only concatenation and reverse traversal operations, i.e., 
{\em simple pattern}.
\ignore{
where inputs from a user is defined in the following grammar: 
\[
p := \; 
\epsilon \bmid a\; (a \bin \mL) \bmid a^-\; (a \bin \mL) 
\bmid p \cdot p
\]
}
Given a simple input pattern, our algorithm leverages the constraints in the 
database to generate a set of {\relexp} patterns {\it related} to 
the input pattern. The algorithm aggregates the similarity scores of these patterns to compute 
the similarity between entities.

\begin{algorithm}[t]
{\small 
\LinesNumbered
\SetKw{KwCont}{continue}
\SetAlgoVlined
\DontPrintSemicolon
\KwIn{schema $S = (\mL,\Gamma)$, 
simple pattern $p = l_1 ... l_n$ over $S$}
\KwOut{subset $\mE_p$ of {\relexp}s over $S$}
\BlankLine
$done \gets \{\}$; \;
$processing \gets \{(\epsilon,0)\}$; \ 
\tcp{\scriptsize For a pair $(r,i) \bin processing$, $r$ is an 
	 {\relexp} $r$ processed up to the position of label $l_i$ in $p$ }
\ForEach{$(r,i) \bin processing$} {
	Remove $(r,i)$ from $processing$; \;
	\If{$i \geq n$}{
		Add $r$ to $done$; \KwCont; \;
	}
	Add $(r \cdot l_{i+1},i+1)$ to $processing$; 
	\label{line:r-exp} \ \tcp{\scriptsize Use original pattern}
	\tcp{\scriptsize Modify each sub-pattern of $s = l_{i+1} \cdot l_{i+2} \cdot ... l_n$}
	\ForEach{$\gamma \bin \Gamma$} {
		$\mR \gets \mR \cup \texttt{ModPatternRefsPerConstraint}(\gamma,s)$; 
		\label{line:callsub} \;
	}
	\ForEach{$j \geq {i+1}$}{
        \ForEach{$(l_{i+1} \cdot l_{i+2} \cdot ... \cdot l_j, e') \bin \mR$} {
			Add $(r \cdot e',j)$ to $processing$; 
            \label{line:r-exp'} \;
		}
	}
}
\KwRet $\mE_p \gets done$; \;
}
\caption{\texttt{PatternGenerator}}
\label{alg:pathmod}
\end{algorithm}
\begin{algorithm}[h]
{\small 
\LinesNumbered
\SetKw{KwCont}{continue}
\SetAlgoVlined
\DontPrintSemicolon
\KwIn{constraint $\gamma$, 
simple pattern $s = l'_1 \cdot ... \cdot l'_m$} 
\KwOut{set $\mR = \{(e,e')\}$ where $e'$ is a corresponding {\relexp} to $e$ which is a sub-pattern of $s$ }
\BlankLine
$G_\gamma \gets$ the premise graph representing $\gamma$; \; 
\ForEach{$i>0, j \geq i, j \leq m$} {
	$e \gets l'_i \cdot l'_{i+1} \cdot ... \cdot l'_j$; \;
	\If {$\mathrm{a\ path\ } e 
		\mathrm{\ from\ some\ } v_g \mathrm{\ to\ } v_h 
		\mathrm{\ exists\ in\ } G_{pre}(\gamma)$} {
		\ForEach{$\mathrm{connected\ subgraph\ } H \mathrm{\ of\ } G_{pre}(\gamma)$} {
			Find all {\relexp}s $e': \btravel{H}{v_g}{v_h}$ 
				that traverse $H$ from $v_g$ to $v_h$ 
				and visit each edge of $H$ once; \;
			Add $(e,e')$ and $(e^-,e'^-)$ to $\mR$; \;
		}
	}
}
\KwRet $\mR$ \;
}
\caption{\texttt{ModPatternRefsPerConstraint}}
\label{alg:subpathmod}
\end{algorithm}

Algorithm~\ref{alg:pathmod} finds
a set $\mE_p$ of {\relexp}s by minimally modifying 
the input simple pattern $p$ 
such that the results of 
Corollary~\ref{crllry:existpathexp} and 
Proposition~\ref{prop:rwrsimrank}
hold for the aggregated scores of all patterns in $\mE_p$.
For example, given an input $p_2:$ 
\texttt{field}$\bcdot$
\texttt{field}$^-$ 
over Figure~\ref{fig:mvd-example2-bib},
the algorithm returns a set of {\relexp}s whose members are including both $p_2$
and $p_4:$
\texttt{field}$\bcdot$
$[\texttt{pub-in}^-]$ $\bcdot$
$[\texttt{pub-in}^-]$ $\bcdot$
\texttt{field}$^-$. 
Our method computes and aggregates the similarity scores of pattern in the set of {\relexp}s returned by Algorithm~\ref{alg:pathmod}
using Equation~\ref{eq:pathsim}.
If $p_5:$ $p_5:$ \texttt{area} $\bcdot$ \texttt{pub\_in} $\bcdot$ $\texttt{pub\_in}^-$ $\bcdot$ $\texttt{area}^-$
is the input of Algorithm~\ref{alg:pathmod} over 
Figure~\ref{fig:mvd-example1-bib}, the algorithm
returns a set of {\relexp}s whose members are
$p_5$ and $p_6: \bskip{\texttt{area} \bcdot \texttt{pub\_in}}$
$\bcdot$ $\bskip{\texttt{pub\_in}^- \bcdot \texttt{area}^-}$.
According to the mapping defined in
Corollary~\ref{crllry:existpathexp} between two data fragments in Figures~\ref{fig:mvd-example2-bib} and 
Figure~\ref{fig:mvd-example1-bib},  
$p_2$ and $p_4$ map to $p_6$ and $p_5$, respectively.
Thus, there is a one-to-one mapping between the sets of {\relexp} 
returned by the algorithm over data fragments.
By computing the scores through counting, 
one finds that the similarity results for $p_2$ over
Figure~\ref{fig:mvd-example2-bib} and $p_5$ over 
Figure~\ref{fig:mvd-example1-bib} are the equal.

More precisely, let Algorithm~\ref{alg:pathmod} take a simple pattern $p = l_1 \bcdot ... \bcdot l_n$ 
over a schema $S = (\mL,\Gamma)$.
Let $(r,i)$ denote a generated {\relexp} $r$ by Algorithm~\ref{alg:pathmod} from $p$, which is processed up to label $l_i$ in $p$.
Given $(r,i)$, let $s$ be the the remaining unprocessed sub-pattern of $p$,  
e.g., $s = l_{i+1} \bcdot ... \bcdot l_n$.
Algorithm~\ref{alg:pathmod} examines each sub-pattern 
$e: l_{i+1} \bcdot l_j$ of $s$ for some $i+1 \leq j \leq n$.
Then, it uses Algorithm~\ref{alg:subpathmod}
to find a set $\mR$ of {\relexp}s for $e$ 
according to each constraint in $S$ (Line~\ref{line:callsub}).
If Algorithm~\ref{alg:subpathmod} generates {\relexp} $e'$ for $e$, 
Algorithm~\ref{alg:pathmod} replaces $e$ in $s$ with $e'$ 
and marks that all labels up to $l_j$ as processed, 
e.g., $(r \cdot e',j)$ (Line~\ref{line:r-exp'}).
It also includes the input pattern $p$ in the set (Line~\ref{line:r-exp}).

Before we explain Algorithm~\ref{alg:subpathmod}, we define the {\it premise graph} of a constraint.
Consider a constraint $\gamma: \phi_\gamma(\bar{x}) \bto \psi_\gamma(\bar{x})$.
We assume that for each atom $(x_i,e,x_j)$ in $\phi_\gamma$,
$e$ is {\it not} written as $e_1 \bcdot e_2$ for some non-empty RPQs $e_1$ and $e_2$.
If such atom exists, we rewrite it as 
$(x_i,e_1,x') \bwedge (x',e_2,x_j)$ for some fresh variable $x'$.
The {\it premise graph} of $\gamma$, denoted by $G_{pre}(\gamma)$, 
is a directed graph $(V,E)$ whose nodes are variables in $\phi_\gamma$ 
and edges are labeled by the RPQ patterns between each pair of those variables in $\phi_\gamma$.
More precisely, a node $v_{x_i} \bin V$ if and only if $x_i$ is a variable in $\phi_\gamma$,
and an edge $(v_{x_i},e,v_{x_j}) \bin E$ if and only if $(x_i,e,x_j)$ is in $\phi_\gamma$.
The premise graph of constraint $\gamma_{1}:$
$(x_1,\texttt{area},x_3) 
\bwedge (x_3,\texttt{pub}\bdashtt\texttt{in},x_4)$ 
$\bwedge (x_2,\texttt{pub}\bdashtt\texttt{in},x_4) 
\bto (x_1,\texttt{area},x_2)$
over Figure~\ref{fig:mvd-example1-bib} is $v_1 \xrightarrow{\text{area}} v_3$
$\xrightarrow{\text{\texttt{pub}\bdashtt\texttt{in}}} v_4$
$\xleftarrow{\text{\texttt{pub}\bdashtt\texttt{in}}} v_2$.

The underlying idea of the Algorithm~\ref{alg:subpathmod} is that
each constraint $\gamma$ implies information\hyp{}preserving transformations 
that may add or remove an edge from the current schema.
These transformations modify only the (simple) patterns in the underlying database
that match the premise of $\gamma$. Intuitively, using the results of Section~\ref{sec:sr-pathsim}, 
these variations are confined within {\relexp}s.
Thus, for each input simple pattern $s$ and constraint $\gamma$, generally speaking, 
Algorithm~\ref{alg:subpathmod} finds all common (sub-)paths between $s$ and the premise
graph of $\gamma$. Let $e$ be such a common path that starts from $v_g$ and ends at $v_h$ in the premise graph of $\gamma$.
Algorithm~\ref{alg:subpathmod} generates all patterns in the premise graph of $\gamma$ that start from $v_g$ and end at $v_h$
and are expressed as {\relexp}s, i.e., all traverses from $v_g$ and end at $v_h$ in the premise graph of $\gamma$ that are expressed as {\relexp}s.

We briefly describe the recursive procedure to 
compute an {\relexp} $\btravel{G}{v_s}{v_t}$ that traverses 
a premise graph $G$ from node $v_s$ to $v_t$ in Algorithm~\ref{alg:subpathmod}.
We adopt a breath\hyp{}first search algorithm
to find all paths, i.e. simple patterns, from node $v_i$ to node $v_j$ in $G$.
An {\relexp} pattern of all $n$ paths that traverses a pair of nodes $v_i$ and $v_j$ 
is $p^{i,j}_1 + ... + p^{i,j}_n$.
Since we assume the constraints and consequently their premise graph to be acyclic, $n$ is exactly 1.
Let us denote this pattern as $p^{i,j}$.
At each node $v_i$ which connects to some leaf node $v_k$ in $G$,
we construct a pattern $[p^{i,k}]$.
Then, we concatenate $[p^{i,k}]$ at the front of any pattern from $v_i$
or at the end of any pattern to $v_i$.
We mark each edge as visited when each pattern $p^{i,j}$ or $[p^{i,k}]$ is constructed.
The base case is to first construct a pattern $p^{s,t}$.
The procedure ends when all edges in $G$ are visited.
We must note that each constructed $p^{i,j}$ can also be written 
as $\bskip{p^{i,j}}$, which results in multiple patterns of this traversal.

As an example, consider the premise graph of constraint $\gamma_1$, $G_{pre}(\gamma_1):$ 
$v_1 \xrightarrow{\text{area}} v_3$
$\xrightarrow{\text{\texttt{pub}\bdashtt\texttt{in}}} v_4$
$\xleftarrow{\text{\texttt{pub}\bdashtt\texttt{in}}} v_2$, 
and a simple pattern $\mathtt{area}\bcdot\mathtt{\texttt{pub}\bdashtt\texttt{in}}$.
Possible {\relexp} patterns that traverse this graph from
$v_1$ to $v_4$, i.e., 
$\btravel{H \subseteq G_{pre}(\gamma)}{v_1}{v_3}$, are
$\mathtt{a} \bcdot \mathtt{p}$, 
$\bskip{\mathtt{a} \bcdot \mathtt{p}}$,
$\mathtt{a} \bcdot \mathtt{p} \bcdot [\mathtt{p^-}]$ and  
$\bskip{\mathtt{a} \bcdot \mathtt{p}} \bcdot [\mathtt{p^-}]$.
Algorithm~\ref{alg:subpathmod} adds all
$(\mathtt{a} \bcdot \mathtt{b}, e')$ to the resulting set $\mR$
where $e'$ is one of the above patterns except the first one.

The complexity of Algorithm~\ref{alg:pathmod} largely
depends on the number of patterns generated in Algorithm~\ref{alg:subpathmod}.
Since each simple pattern $p$ found in $G_\gamma$ can be either $p$ or $\bskip{p}$, 
this algorithm is exponential in the number of $p$'s.
However, since $G_\gamma$ is acyclic, each $p$ is a path. 
Hence, the number of simple pattern $p$'s is 
$1+\sum_{v \bin V_{\deg>2}} (\deg(v)-1)$
where $V_{\deg>2}$ is a set of nodes in $G$ whose degrees are greater than 2.
Since $\sum_{v \bin V(G)} \deg(v) = 2|E(G)|$,
the procedure is $O(\mathrm{exp}(|E(G)|))$.
There is also a linear\hyp{}time algorithm $O(V(G)$
that finds all connected subgraphs $G$ of $G_\gamma$ \cite{Hopcroft:1973:AEA:362248.362272}.
Since the number of iterations over all sub\hyp{}patterns 
(Line 2) is polynomial in the length of $p$, 
the total complexity of Algorithm~\ref{alg:subpathmod} is
$O((n^2)(\mathrm{exp}(|E(G)|)+|V(G)|))$ where $n$ is the number of nodes in the
input simple pattern.
Constraints of a schema are usually 
have a very small number of terms compared to the size of databases. 
We may view the exponent factor as a (small) constant.
Hence, Algorithm~\ref{alg:subpathmod} is $O(n^2)$ where $n$ is usually small \cite{Sun:VLDB:11}.
Let $O(M)$ denote the number of patterns
generated from Algorithm~\ref{alg:subpathmod}.
Consider that there are possible $O(\mathrm{exp}(k))$ 
possible sub\hyp{}patterns to the user input pattern with $k$ nodes.
Also, algorithm~\ref{alg:pathmod} may replace each of those sub\hyp{}patterns by the generated patterns from  
Algorithm~\ref{alg:subpathmod}.
Hence, the complexity of Algorithm~\ref{alg:pathmod} is $O(M^{\mathrm{exp}(k)})$.
The value of $k$, i.e., the number of nodes in the submitted simple pattern is also significantly smaller than the size of the database.
We will discuss the optimizations to improve the running time of Algorithm~\ref{alg:subpathmod} in Section~\ref{sec:robustalg-simple-alg}.

Consider each label $l$ in the input simple pattern.
Algorithm~\ref{alg:subpathmod} finds all possible
patterns according to the set of constraints 
that is mapped to each label $l$ in the input pattern and follows Theorem~\ref{thrm:existpathexp}.
Using similar arguments to Theorem~\ref{thrm:existpathexp}
and Corollary~\ref{crllry:existpathexp}, 
we have the following proposition.
\begin{proposition}\label{prop:aggrelsim}
Given a database $D$ of schema $S$ and a equivalent schema $T$ 
under transformation $\Sigma$, 
for every simple pattern $p_S$ over $S$, there exists 
a simple pattern $p_T$ over $T$ such that, $\forall u,v \bin D$, 
$\sum_{p \bin \mE_{p_S}} \texttt{sim}_p(u,v,D) =$ $
\sum_{p' \bin \mE_{p_T}} \texttt{sim}_{p'}(u,v,\Sigma(D))$
\end{proposition}

\begin{proof}
Let $f_{crpq}(\btravel{G}{v_g}{v_h})$ denote a CRPQ 
representing an {\relexp} $exp': \btravel{G}{v_g}{v_h}$
in Algorithm~\ref{alg:subpathmod}.
Consider a transformation 
$\gamma: f_{crpq}(\btravel{G}{v_g}{v_h}) \bto (x_g,l,x_h)$ 
for some variables $x_g$ and $x_h$ corresponding to $v_g$ and $v_h$, respectively,
and some label $l \notin S$.
Clearly, $\gamma$ is information preserving, 
and $\bskip{\btravel{G}{v_g}{v_h}}$ is mapped to $l$
according to Theorem~\ref{thrm:existpathexp}.
Similarly, there exists an {\relexp} $exp''$ over $T$ that 
maps to $l$ for some $\Sigma'$.
By transitivity, $exp'$ is mapped to $exp''$
according to Theorem~\ref{thrm:existpathexp}.
We must note that $\gamma$ always exists, and 
Algorithm~\ref{alg:subpathmod} yields all possible such $\Sigma'$'s.
Hence, using similar arguments to 
Theorem~\ref{thrm:existpathexp}
and Corollary~\ref{crllry:existpathexp}, 
we have that our proposition holds.
\end{proof}

\noindent
Thus, a similarity algorithm that computes 
aggregate scores over the set of {\relexp}s 
returned by Algorithm~\ref{alg:pathmod} is structurally robust.

\section{Optimizing Pattern Generation}
\label{sec:robustalg-simple-alg}
We have simplified the use of our framework in Section~\ref{sec:robustalg-simple}
so that users can take advantage of it by submitting only simple patterns.
Using database constraints, our algorithm finds a set of relationship patterns related to the input pattern and use them to compute 
an aggregated similarity score.
If the set of such {\relexp}s patterns is too large, our system has to compute the similarity scores for many patterns, therefore, 
it may {\it not} be efficient on a large database.
In this section, we show that the algorithms proposed in Section~\ref{sec:robustalg-simple} can avoid generating and computing the similarity scores 
of many such patterns without losing their robustness or effectiveness.

According to our discussion in Section 3, to have an invertible variation of a database $I$, $I$ must satisfy some tgd constraints. 
However, these constraints may be {\it trivial} , e.g., $(x,a,y) \bto (x,a,y)$. 
Obviously, it is {\it not} efficient to consider all trivial constraints over a database in the algorithms proposed to simplify our system. 
We show that a structural variation actually needs $I$ to have non-trivial constrains. 
Thus, as far as our proposed algorithms use non-trivial constraints, it is structurally robust. 
We also show that a simple pattern will be restructured based on a tgd only if at least one of its labels appear in both left- and right hand-side of the tgd. 
This enables Algorithm~\ref{alg:pathmod} to ignore many tgds for a given input pattern. 
These optimizations reduce the number of patterns need to be generated by Algorithm 1 and significantly improve its running time and the running time of its similarity search.

\subsection{Avoiding Trivial Constraints}
Our proposed Algorithms~\ref{alg:pathmod} and~\ref{alg:subpathmod}
indeed require a non\hyp{}empty set of a schema constraints.
According to Proposition~\ref{theorem:SourceConstraints}, 
structural variations require database constraints.
However, the proposition does not exclude the use of trivial constraints.
Intuitively, a trivial constraint is a constraint such that 
its premise and its conclusion are logically equivalent,
e.g., $\phi(\bar{x}) \bto \phi(\bar{x})$.
Clearly, every database in a schema $S$ satisfies 
all trivial constraints for every label $a \bin S$. 
In order to simplify our discussion, since trivial constraints do {\it not} put any restriction on a database, 
if a schema (database) satisfies only trivial constraints, we say that it does {\it not} satisfy any constraint.

In relational database, it has been proved by Hull that
there is {\em not} any variation of a relational schema with the same information without any constraint 
beyond simple renaming of the scheme elements \cite{infopreserve:hull}.
We show that this result also holds for graph databases. Given transformation $\Sigma_{ST}$ and $I \in \inst{S}$, 
let $img_{\Sigma_{ST}}(I)$ be the set of databases where each is created by removing all nodes and edges constructed using existentially quantified variables in $\Sigma_{ST}$ in each $J \in \Sigma_{ST}$. We denote the set of $img_{\Sigma_{ST}}(I)$ for all $I \in \inst{S}$ as $img_{\Sigma_{ST}}(S)$

\begin{theorem}
\label{thrm:variations-constaint}
Given two schemas $S = (\mL_S, \Gamma_S)$ and $T = (\mL_T, \Gamma_T)$
where $\Gamma_S$ and $\Gamma_T$ are empty,  
if there exists an invertible $\Sigma_{ST}$ between $S$ and $T$
there is a bijection between $\mL_S$ and the labels used in $img_{\Sigma_{ST}}(S)$.
\end{theorem}
\begin{proof}
Suppose $S \neq T$ and there is no bijection between $\mL_S$ and $\mL_T$.
Hence, $|S| \neq |T|$.
Without losing generality, assume $|S| > |T|$.
For a given set of nodes $V$, we have that the set 
$V \times \mL_S \times V$ has the order of 
$|V| \times |\mL_S| \times |V| > |V| \times |\mL_T| \times |V|$
in which the latter is the order of $V \times \mL_T \times V$.
Let $\instv{S}{V}$ denote a set of database instances of $S$ whose
vertex set is $V$.
By the definition of a database, for each $D_S \bin \instv{S}{V}$, 
$E_{D_S} \subseteq V \times \mL_S \times V$.
Similarly, for each $D \bin \instv{T}{V}$, 
$E_{D_T} \subseteq V \times \mL_T \times V$.
Without any restriction in the set $E_{D_S}$ and $E_{D_T}$, 
we have that $|\instv{S}{V}|>|\instv{T}{V}|$.
That is, for any surjective mapping function from 
$\instv{S}{V}$ to $\instv{T}{V}$, there exists an instance of $\instv{S}{V}$ that maps to multiple instances of $\instv{T}{V}$.
Since the statement holds for any $V \subseteq \mV$, 
we have that there is no surjective mapping function from 
$\instv{S}{V}$ to $\instv{T}{V}$ that is also injective.
Hence, there exists no bijection between $\inst{S}$ and $\inst{T}$.
Therefore, no such $\Sigma_{ST}$ exists.
\end{proof}

\noindent
Thus, we have that for a representation variations 
beyond renaming, either the source schema or the target schema
must contain some non-trivial constraints.

Assume that all associated constraints with a schema $S$ are trivial.
It is possible to have a non\hyp{}identity transformation $\Sigma_{ST}$ 
from $S$ to a target schema $T$.
For instance, consider a schema $S = \{a,b\}$ without any non-trival constraint.
A transformation $\Sigma_{ST}$ from $S$ to a target schema $T = \{a,b,c\}$ 
described as
$\{ (x_1,a,x_2) \bwedge (x_2,b,x_3) \bto (x_1,c,x_3)$ and 
$(x_1,l,x_2) \bto (x_1,l,x_2), l \bin S \}$
is invertible.
In this case, $T$ consists of a constraint
$(x_1,a,x_2) \bwedge (x_2,b,x_3) \bto (x_1,c,x_3)$.
However, with only trivial constraints available in $S$, 
Algorithm~\ref{alg:subpathmod} does {\it not} produce
an {\relexp} $\bskip{a \bcdot b}$ over $S$ which is mapped to $c$ over $T$
because the pattern involves two separate constraints.
In fact, with only a trivial constraint, the algorithms does not modify an input pattern.

\ignore{
In addition, following Proposition~\ref{theorem:SourceConstraints}, 
we define that {\bf a transformation induced by 
a constraint $\gamma$ over schema $S$} is
{\bf an information preserving transformation $\Sigma_{ST}$ 
whose inverse $\Sigma_{ST}^{-1}$ satisfies
$\Sigma_{ST}^{-1} \circ \Sigma_{ST} \equiv \gamma$}.
}

Using the result of Proposition~\ref{theorem:SourceConstraints}, 
we have the following theorem regarding constraints of $T$.

\begin{theorem}\label{thrm:constequivnone}
Given two schemas $S = (\mL_S, \Gamma_S)$ and $T = (\mL_T, \Gamma_T)$
where $\Gamma_T = \emptyset$,  
if there is an invertible transformation $\Sigma_{ST}$, 
then there is a bijection between $L' \subseteq \mL_{S}$ and $\mL_T$ 
where there is no constraint in $S$ whose conclusion contains label in $L'$, 
and for each $l \bin \mL_S \setminus L'$, there exists a constraint 
$\lambda(\bar{x}) \bto (x_1,l,x_2)$ in $\Gamma_S$ where $l$ does not appear in any atom in $\lambda$.
\end{theorem}
\begin{proof}
It is implied by Theorem~\ref{thrm:variations-constaint} that 
$|\mL_S| \leq |\mL_T|$.
Further, using similar argument in the proof of Theorem~\ref{thrm:variations-constaint},
we have that, if there exists no $L' \subseteq \mL_T$ such that a bijection
between $\mL_S$ and $L'$ exists, then there is no bijection between $\inst{S}$ and $\inst{T}$.

Consider a set of instances $V \subseteq \mV$.
We have that, since there is no constraints whose conclusion contains a label in $L'$, 
for each $I \bin V \times L' \times V$, there exists $I' \bin \inst{T}{V}$ and $I \subseteq I'$.
Suppose there is no constraint $\lambda(\bar{x}) \bto (x_1,l,x_2)$, 
for some $l \bin \mL_T \setminus L'$.
For some subset $V \subseteq \mV$, 
there must exist a database $J$ containing edges of label $\mL_T \setminus L'$
and there is no $K \bin V \times L' \times V$ s.t. $K \subseteq J$.
Therefore, $|\instv{T}{V}| > |V \times L' \times V| \bequal |\instv{S}{V}|$.
That is, a surjective mapping between $\inst{S}{V}$ and $\instv{T}{V}$, 
there exists an instance in $\instv{S}{V}$ that maps to multiple instances in $\instv{T}{V}$.
In addition, since the bijection between $\mL_S$ and $L'$ exists, 
for any $V'\neq V \subseteq \mV$, $|\instv{S}{V'}| \leq |\instv{T}{V'}|$.
Using similar argument to Theorem~\ref{thrm:variations-constaint}, 
we show that there is no bijective mapping between $\inst{S}$ and $\inst{T}$.
Therefore, $S$ and $T$ are not information equivalent.
\end{proof}

Following Theorem~\ref{thrm:constequivnone}, 
we have that if a schema $S$ contains no constraint, 
every invertible transformation from $S$ preserves all edges 
(or up-to renaming of those edges). We call this type of transformations {\em easy}.

Based on Theorem~\ref{thrm:constequivnone},  
our proposed algorithms can ignore producing any pattern over a non-trivial constraint
in the form of $\phi(\bar{x}) \bto (x_1,l,x_2)$
where $l$ does {\it not} appear in a CRPQ $\phi$.
Consider a schema $S$ whose constraint 
is $\phi(\bar{x}) \bto (x_1,l,x_2)$, where $l$ does not appear in $\phi$. 
There is an invertible transformation from $S$ to a schema $T = S \setminus \{l\}$.
We have that $\btravel{G_\phi}{x_1}{x_2}$ exists in both $S$ and $T$.
Also, following the proof of Theorem~\ref{thrm:existpathexp},
we have that pattern $l$ over $S$ is mapped to a pattern $r: \bskip{\btravel{G_\phi}{x_1}{x_2}}$.
However, $r$ is not a simple pattern and might not be easily discovered by a user.
If we would like to ensure the robustness of {\rpathsim}
via the use of Algorithm~\ref{alg:pathmod}, 
then either label $l$ should be ignored
or every $l$ should be replaced with $\btravel{G_\phi}{x_1}{x_2}$
that does {\it not} contain a skip-operation.

\subsection{Filtering Constraints Based on Their Conclusions}
Intuitively, in order to modify a database structure, 
a transformation may add or remove edges of certain labels.
Also, we have shown that a database constraint, 
either in source or target schema, 
is necessary for structural variations beyond renaming.
However, not every label appearing in the constraint can be removed.
Following Proposition~\ref{theorem:SourceConstraints},
let us define a transformation induced by 
a constraint $\gamma$ over schema $S$, denoted by $\Sigma_{ST}^{\gamma}$, 
as an invertible transformation $\Sigma_{ST}$ 
whose inverse $\Sigma_{ST}^{-1}$ satisfies
$\Sigma_{ST}^{-1} \circ \Sigma_{ST} \equiv \gamma$.
We show in Proposition~\ref{prop:non-removal-label} 
that, given a constraint $\gamma$, 
an invertible transformation induced by $\gamma$ may remove only 
edges of a label that appears in the conclusion of $\gamma$.

\begin{proposition}\label{prop:non-removal-label}
Given a schema $S$ with a constraint 
$\gamma: \phi_\gamma(\bar{x}) \bto (x_1,a,x_2)$,
for every invertible transformation $\Sigma_{ST}^{\gamma}$
from $S$ to a target schema $T$,
there exists a mapping $M: \Sl{S} \bto \Sl{T}$ such that 
$(x,l,y) \bto (x,M(l),y)$, for all $l \neq a \bin \Sl{S}$,
in $\Sigma_{ST}^{\gamma}$.
\end{proposition}
\begin{proof}
Let $r \neq a$ be a label in schema $S$.
Consider $D_1, D_2 \bin \inst{S}$ in which 
$D_1$ consists of nodes $u$, $v$ and an edge $(u,r,v)$
and $D_2$ consists of nodes $u$ and $v$ without $(u,r,v)$.
Suppose $(x,r,y) \bto (x,M_l(r),y)$ does not exists
in $\Sigma_{ST}^{\gamma}$.
Then, we have that there is no edge between $u$ and $v$ in 
$\Sigma_{ST}^{\gamma}(D_1)$.
That is, $\Sigma_{ST}^{\gamma}(D_1) = \Sigma_{ST}^{\gamma}(D_2)$.
Since $\Sigma_{ST}^{\gamma}$ is information preserving, 
then $D_1 = D_2$ which is contradiction.
Hence, the proposition holds.
\end{proof}

Consider that each transformation rule
$(x,l,y) \bto (x,M(l),y)$ simply renames each edge label
$l$ to a new edge label $M(l)$ in the target schema.
For simplicity of our model and analysis, 
we refer to $M(l)$ in the target schema simply as $l$ 
and assume that this transformation rule always exists.
\texttt{published\hyp{}in} is an example of such label 
in the transformation between the two databases presented 
in Figure~\ref{fig:mvd-example-bib}.
We also call a transformation that preserves all edges
that appears in the premise of a constraint an {\em easy} transformation. 

Some constraint may induce an invertible transformation that is {\it not} easy.
For instance, a transformation between structure of databases
shown in Figure~\ref{fig:mvd-example-bib} is {\it not} easy. 
However, not every non\hyp{}easy transformation $\Sigma$
is invertible unless there exists an inverse $\Sigma^{-1}$
such that $\Sigma^{-1} \circ \Sigma$ is equivalent to the constraint.
In this paper, we do {\it not} provide a procedure to determine 
whether a transformation is invertible.
Regardless, using Propositions~\ref{prop:non-removal-label}, 
we may conclude that non\hyp{}easy transformations
are induced by some constraint 
$\phi(\bar{x}) \bto (x_1,l,x_2)$
where $l$ appears in $\phi(\bar{x})$.
That is, an {\relexp} that does {\it not} contain label $l$
is obtained from some easy transformation.

To this end, we should filter out all 
{\relexp}s returned by Algorithm~\ref{alg:pathmod} 
that are induced by any easy transformation. 
For instance, given a constraint 
$(x_1,\texttt{area},x_3) 
\bwedge (x_3,\texttt{published-in},x_4)
\bwedge (x_2,\texttt{published-in},x_4) 
\bto (x_1,\texttt{area},x_2)$
in Figure~\ref{fig:mvd-example1-bib}, 
the algorithm ignores producing an {\relexp} such as\\
\texttt{published-in}$\bcdot$\texttt{published-in}$^-$.
However, an {\relexp} such as 
\texttt{area}$\bcdot$\texttt{published-in}
is valid because \texttt{area} appears in the conclusion
of the constraint.
Hence, this filtering helps reduce the space and running time of aggregate
{\rpathsim} over a set of relationship patterns
returned by Algorithm~\ref{alg:pathmod}.

\section{Empirical Evaluation}
\label{sec:experiment}

\newcommand{\biomed}{BioMed}

\begin{figure}[t]
\centering
\subfigure[DBLP]{
\includegraphics[width=0.21\textwidth]{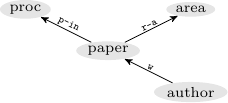} 
\label{fig:dblp1}}
\subfigure[SIGMOD Record]{
\centering
\includegraphics[width=0.21\textwidth]{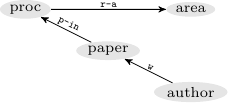}
\label{fig:dblp2}}
\caption{Schema fragments of bibliographic databases.
\texttt{p-in}, \texttt{r-a} and \texttt{w} denote edge labels
\texttt{published}\hyp{}\texttt{in}, 
\texttt{research}\hyp{}\texttt{area} and
\texttt{writes}, respectively.}
\label{fig:dblp}
\end{figure}

\begin{figure}[t]
\centering
\subfigure[WSU]{
\includegraphics[width=0.21\textwidth]{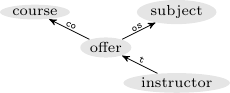} 
\label{fig:course1}}
\subfigure[Alchemy UW-CSE]{
\centering
\includegraphics[width=0.21\textwidth]{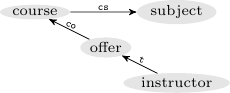}
\label{fig:course2}}
\caption{Variations for course databases.
\texttt{cs}, \texttt{os}, \texttt{t} and \texttt{co} 
denote edge labels
\texttt{course}\hyp{}\texttt{subject}, 
\texttt{offering}\hyp{}\texttt{subject}, 
\texttt{teach}, and
\texttt{offering}\hyp{}\texttt{course}, respectively.}
\label{fig:course}
\end{figure}

\begin{figure}[t]
\centering
\includegraphics[width=0.48\textwidth]{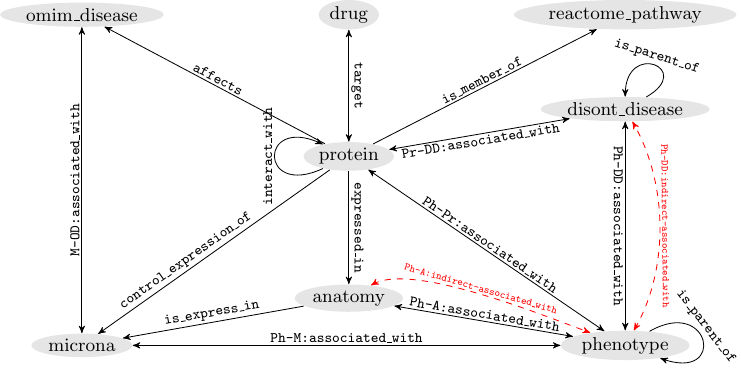}
\caption{Schema fragment of {\biomed} dataset}
\label{fig:biomed}
\vspace{-15pt}
\end{figure}

\noindent {\bf Datasets:}
We use 4 datasets in our experiments:
DBLP, Microsoft Academic Search (MAS), WSU course dataset, and a Biomedical dataset ({\biomed}).
{\em DBLP} consists of 1,227,602 nodes and 2,692,679 edges, 
which contains bibliographic information of 
publications in computer science.
We add information about the research areas
for each conference in DBLP from information 
extracted from Microsoft Academic Search (MAS).
Figure~\ref{fig:dblp1} depicts the schema of DBLP.
We also use a subset of Microsoft Academic Search data with 44,068 nodes and 44,220 edges.
MAS contain information about papers, conferences, areas, e.g., {\it Databases}, and keywords of each paper and/or area, e.g., {\it indexing}. 
WSU course database\footnote{\it cs.washington.edu$\bslash$research$\bslash$xmldatasets}
contains information about courses, instructors and course offerings in the university.
The dataset consists of 1,124 nodes and 1,959 edges.
Figure~\ref{fig:course1} depicts the schema of WSU dataset.
The Biomedical dataset ({\biomed}), is made available to us as a part of an NIH funded project 
and contains information about genetic conditions, diseases, drugs, and their relationships. 
Figure~\ref{fig:biomed} depicts a fragment of {\biomed}.
It consists of 43,307 nodes and 1,742,970 edges. 

\noindent {\bf Settings:} We compare robustness, effectiveness
and efficiency of {\rpathsim} with
RWR \cite{Tong:ICDM:06} using a restart probability of 0.8,
SimRank \cite{Jeh:KDD:02} using a damping factor of 0.8,  
and PathSim \cite{Sun:VLDB:11}. 
Since previous studies show that the PathSim similarity computation method 
is more effective than those of SimRank and RWR \cite{Sun:VLDB:11}, 
we use {\rpathsim} with the similarity score computation of PathSim, i.e., Equation~\ref{eq:pathsim}.
We implement all algorithms using MATLAB 8.5
on a Linux server with 64GB memory 
and two quad-core processors.

\noindent 
{\bf Result Overview:} Our empirical study indicates that current algorithms are {\it not} robust over invertible transformations. 
Moreover, our proposed algorithm are also generally more robust than other algorithms over transformations in which the original database may have less information than the transformed one. This indicates that our algorithm may also tolerate losing some information during restructuring of the data better than other algorithms. Our results also indicate that our algorithms returns at least as effective or more effective results than other algorithms over real-world datasets and query workloads. Finally, we show that although RelSim uses a more expressive relationship language than PathSim, it is as efficient as PathSim and runs efficiently over large databases. We have also evaluated the efficiency of RelSim that uses simple input patterns as explained in Section 5. Our results over a large real-world database and query workload with varying number of constraints on the database and the length of input patterns suggest that RelSim is efficient over a large database with relatively large number of constraints and long input patterns.


\subsection{Structural Robustness}
\label{sec:exp-robust}
We use DBLP, WSU and {\biomed} databases to
evaluate the structural robustness of
RWR, SimRank, PathSim and {\rpathsim}.
As SimRank takes too long to finish on full DBLP dataset, 
we do this evaluation using 
a subset of DBLP with 24,396 nodes and 98,731 edges.

DBLP dataset satisfies constraint
$(paper_1, \texttt{r-a}, area) 
\bwedge (paper_1, \texttt{p-in}, proc) 
\bwedge (paper_2, \texttt{p-in}, proc)
\bto (paper_2, \texttt{r-a}, area)$.
We transform this database  
to a database with the structure shown in Figure~\ref{fig:dblp2}, which follows the style of SIGMOD Record database.
We call this transformation {\sffamily DBLP2SIGM}.
We randomly sample 100 proceedings based on their node degrees as our query workload over these datasets.

WSU course dataset satisfies the constraint
$(offer_1, \texttt{os}, subject) 
\bwedge (offer_1, \texttt{co}, course) 
\bwedge (offer_2, \texttt{co}, course)$ \\ 
$\bto (offer_2, \texttt{os}, subject)$. 
We transform WSU course database to the graph structure of 
Alchemy UW\hyp{}CSE database\footnote{\it alchemy.cs.washington.edu$\bslash$data$\bslash$uw-cse}
whose structure is shown in Figure~\ref{fig:course2}.
We call this transformation {\sffamily WSUC2ALCH}.
We also randomly sample 100 courses from WSU based on their degrees as our query workload for these datasets.

{\biomed} dataset satisfies 
({\it phenotype}$_1$, \texttt{is}$\bdashtt$\texttt{parent}$\bdashtt$\texttt{of}, \\ {\it phenotype}$_2$) 
$\bwedge$ 
 ({\it phenotype}$_1$, \texttt{associated}$\bdashtt$\texttt{with}, {\it anatomy})
 $\bto$
({\it phenotype}$_2$, \texttt{indirect}$\bdashtt$\texttt{associated}$\bdashtt$\texttt{with}, {\it anatomy})
and
({\it phenotype}$_1$, \texttt{is}$\bdashtt$\texttt{parent}$\bdashtt$\texttt{of}, {\it phenotype}$_2$) 
$\bwedge$ 
({\it disease}, \texttt{associ}- \texttt{ated}$\bdashtt$\texttt{with}, {\it phenotype}$_1$) $\bto$
({\it disease}, \texttt{indirect}$\bdashtt$\texttt{associated}$\bdashtt$\texttt{with}, {\it phenotype}$_2$).
We transform the {\biomed} dataset such that 
all edges of label 
\texttt{indirect}\hyp{}\texttt{associated}\hyp{}\texttt{with}
are removed.
We denote the transformation over {\biomed} dataset 
as {\sffamily {\biomed}T}.
The structure of the transformed {\biomed} dataset is also shown in Figure~\ref{fig:biomed} excluding all dashed edges.
The main goal of using this dataset in the NIH project is to discover the drugs that are closely related to queried diseases.
Since we use this dataset to also evaluate the effectiveness of our algorithms, 
we have obtained a set of 30 diseases and their relevant drugs from experts in the domain of the data.
Since paths between diseases and drugs are asymmetric, 
we cannot compute similarity scores using PathSim formula 
over this dataset.
Instead, we evaluate the queries using HeteSim
\cite{shi2014hetesim}, 
which extends PathSim to support asymmetric paths, 
e.g., finding similarity between different entity types.

We measure the structural robustness of each method by comparing its ranked list of results for the same query 
over different datasets with the same information but different structural representations.
We adopt normalized Kendall's tau measurement 
to compare two ranked lists.
The value of normalized Kendall's tau varies
between 0 and 1 
where 0 means two lists are identical
and 1 means one list is the reverse of another.
As users are normally interested in highly ranked answers, 
we compare top 5 and 10 answers.

Table~\ref{table:robustness} shows the average 
ranking differences for top 5 and 10 answers
returned by RWR, SimRank, PathSim {\andor} HeteSim.
We do {\it not} report the results of {\rpathsim}
because it returns the same answers over all transformations.
According to Table~\ref{table:robustness},
the outputs of all current algorithms
are significantly different across databases under 
these information-preserving transformations.

\begin{table}[t]
\caption{Average ranking differences}
\vspace{-10pt}
\centering
\begin{tabular}{rc@{\hspace{1em}}cc@{\hspace{1em}}cc@{\hspace{1em}}c}
\cline{2-7}
\multicolumn{1}{l}{} & \multicolumn{2}{c}{{\sffamily DBLP2SIGM}} & \multicolumn{2}{c}{{\sffamily WSUC2ALCH}} & \multicolumn{2}{c}{{\sffamily {\biomed}T}} \\
\cline{2-7}
\multicolumn{1}{l}{} & top 5 & top 10 & top 5 & top 10 & top 5 & top 10 \\
\hline
RWR 		& .447	& .412	& .259	& .253	& .130	& .112	\\
SimRank		& .455	& .410	& .387	& .341	& .405	& .385	\\
PathSim		& .608	& .590	& .310	& .247 	& .438	& .461 	\\
\hline
\end{tabular}
\vspace{-10pt}
\label{table:robustness}
\end{table}

We also evaluate the robustness of our algorithm 
when a transformation is invertible and adds extra information.
We define a transformation, called {\sffamily DBLP2SIGMX},
similar to {\sffamily DBLP2SIGM}, which also adds additional nodes that connect authors to proceedings in the SIGMOD Record schema. These nodes indicate the relationship of what authors has a paper in what proceedings. Each of these nodes is connected to both the corresponding author and the proceedings in wich she has published at leas one paper. {\sffamily DBLP2SIGMX} is invertible and has the same inverse as {\sffamily DBLP2SIGM}.
It takes more than a day to run SimRank and RWR over DBLP and {\biomed}.
Thus, to compare the impact of adding information on all algorithms, 
we have performed these experiments over the small DBLP dataset, 
and a subset of {\biomed} datasets which has 4,125 nodes and 60,176 edges. 
We randomly selected 100 queries for DBLP dataset and use the same query workload used in Section 6.1 for {\biomed} dataset.
We also use the relationship patterns for PathSim, HeteSim and {\rpathsim} that are used in the previous robustness experiments.
Table~\ref{table:robustness-extra} shows the robustness of all algorithms, including RelSim, under this transformation, which indicate that RelSim is the only robust algorithm over this transformation.

Furthermore, some real-world data transformations may {\it not} be invertible  and lose some information.
Hence, we also measure the robustness of algorithms 
when a small fraction of edges between nodes are removed during the transformation.
In this experiment, we create two new transformations for DBLP and BioMed, 
namely {\sffamily DBLP2SIGM(.95)} and {\sffamily BioMedT(.95)}, respectively.
{\sffamily DBLP2SIGM(.95)} and {\sffamily BioMedT (.95)} 
restructure DBLP and {\biomed} similar to that of 
{\sffamily DBLP- 2SIGM} and {\sffamily BioMedT}, respectively.
These transformation first restructure the database and then randomly remove 5\% of the total number of edges from the transformed databases.
Table~\ref{table:robustness-extra} shows the average ranking differences 
for top 5 and top 10 answers returned by RWR, SimRank, PathSim (HeteSim) and {\rpathsim} using the same datasets, 
the query workloads and relationship patterns described in the previous experiment. The result indicates that {\rpathsim} is generally more robust than other methods when the transformed databases has a slightly less information than the original ones. 

\begin{table}[t]
\caption{Average ranking differences over transformations that modify information}
\vspace{-10pt}
\centering
{\small
\begin{tabular}{rc@{\hspace{1em}}cc@{\hspace{1em}}cc@{\hspace{1em}}c}
\cline{2-7}
\multicolumn{1}{l}{} & \multicolumn{2}{c}{{\sffamily DBLP2SIGM}} &  \multicolumn{2}{c}{{\sffamily {\biomed}T(.95)}} & \multicolumn{2}{c}{{\sffamily DBLP2SIGM(.95)}} \\
\cline{2-7}
\multicolumn{1}{l}{} & top 5 & top 10 & top 5 & top 10 & top 5 & top 10 \\
\hline
{\rpathsim}         & 0  & 0  & .750  & .144 & 0 .170  & .298 \\
RWR 		        & .696  & .752 & .530  & .500 & .835 & .640 \\
SimRank		        & .790	& .750	& .143	& .344 & .790 & .750 \\
PathSim	        	& .364	& .239	& .927	& .927 &  .423 & .452 \\
\hline
\end{tabular}}
\vspace{-10pt}
\label{table:robustness-extra}
\end{table}


\subsection{Effectiveness}
\label{sec:exp-effectiveness}

We evaluate the effectiveness of {\relexp} over {\biomed} and MAS databases. We use the same query workload used in Section 6.1 for {\biomed}
Since each disease query relates only to a single drug, 
we use {\it Mean Reciprocal Rank} (MRR) to evaluate the effectiveness of the algorithms.
{\it Reciprocal Rank} (RR) of a list of answers to a query is $1 / p$ where $p$ is the position of the first relevant answer in the returned list of answers. MRR is the average of RR over a set of queries.

\begin{table}[t]
\caption{{Average MRR of algorithms over {\biomed}.}}
\vspace{-10pt}
\label{table:biomed-effectiveness}
\centering
\begin{tabular}{rc@{\hspace{1em}}c@{\hspace{1em}}c@{\hspace{1em}}c}
\hline
{\biomed} dataset & RWR & SimRank & HeteSim & {\rpathsim} \\
\hline
original      
& .010  & .062  & .077  & .077 \\
under {\sffamily {\biomed}T} 
& .010  & .062  & .072  & .077 \\
\hline
\end{tabular}
\vspace{-10pt}
\end{table}

Table~\ref{table:biomed-effectiveness} 
shows average MRR of RWR, SimRank, HeteSim
and {\rpathsim} over original {\biomed} dataset 
and {\biomed} under {\sffamily {\biomed}T}.
According to our discussion with the experts, these queries are very hard to answer effectively by using only the structural patterns in the data set and without consulting external sources of knowledge and even a slight improvement in the accuracy of the returned answers may save a great deal of time and effort in their research.  The overall results show that {\rpathsim} are more effective than other algorithms.
This implies that the use of {\relexp} language helps to improve the effectiveness of the algorithm.
We have also used the BioMed and query workload to evaluate the effectiveness of similarity search method proposed in Section 5. We have used the simple pattern given to HeteSim for this experiment. The MRR of the final results is close to the one delivered by RelSim algorithm that takes RRE patterns as input, which is as effective or more effective than other algorithms. 
These results also indicate that RelSim is as effective or more effective than similar algorithms.


\subsection{Efficiency}
\label{sec:exp-efficiency}
We evaluate the query processing time of {\rpathsim} and PathSim over DBLP and {\biomed} datasets using the query workloads reported in Section~\ref{sec:exp-robust}.
As explained in Section 6.1, it takes almost a day to run SimRank and RWR over these datasets.
First, we evaluate the query processing time of {\rpathsim} and PathSim for the case where the user provides an exact relationship pattern (Section \ref{sec:robustalg}).
All reported running times in this section assume that 
the commuting matrices of all meta-paths, i.e., simple {\relexp} patterns that use only concatenation and reversal operations, up to size 3 
are materialized and pre-loaded in main memory for both {\rpathsim} and PathSim.
Theoretically, both {\rpathsim} and PathSim have the same time complexity. 
However, the expressiveness of {\relexp} used in 
{\rpathsim} allows the specified relationship pattern to be 
more complex than the expression used by PathSim.
To compare the efficiency between these two algorithms,
we first pick a pattern over each database as a reference.
Then, for each referenced pattern, 
we find the corresponding {\relexp} pattern $p_R$ for {\rpathsim}
and the closest correspondent simple pattern, i.e., meta-path, $p_P$ for PathSim. 
For instance, a referenced pattern over DBLP is 
\texttt{p-in}$\cdot$\texttt{r-a}.
Over DBLP under {\sffamily DBLP2SIGM} transformation, 
the correspondent patterns for {\rpathsim} and PathSim are 
$p_R: [\texttt{p-in}^-]\cdot$\texttt{r-a}
and $p_P: \texttt{r-a}$, respectively.
Then we compare the running time of PathSim using $p_P$
with the running time of {\rpathsim} using $p_R$ and report the results.
\begin{table}
\caption{ Average query processing time in seconds.}
\vspace{-10pt}
\label{table:simple-efficiency}
\begin{tabular}{rcc@{\hspace{2em}}cc}
\cline{2-5}
& \multicolumn{2}{c}{single pattern} & \multicolumn{2}{c}{using Algorithm~\ref{alg:pathmod}} \\
\cline{2-5}
& DBLP & {\biomed} & DBLP & {\biomed} \\
\hline
{\rpathsim} & .035  & .473 & .034   & .511 \\
PathSim     & .024  & .267 & .027   & .477 \\
\hline
\end{tabular}
\vspace{-10pt}
\end{table}
\ignore{
The average query processing time for a single pattern per query 
of {\rpathsim} (PathSim) over DBLP and {\biomed} dataset are 
0.035 (0.024) and 0.473 (0.267) seconds, respectively.
}
Table~\ref{table:simple-efficiency} show the average query processing time 
for a single pattern per query of {\rpathsim} and PathSim (under "single pattern").
Overall, {\rpathsim} is slower than PathSim because {\rpathsim} uses
more complex and longer patterns than those used by PathSim. 
Nevertheless, the running time of {\rpathsim} is still relatively short 
over large datasets.

Next, we measure the efficiency of {\rpathsim} that incorporates
Algorithm~\ref{alg:pathmod} introduced in Section~\ref{sec:robustalg-simple} with the the optimization techniques to ignore non-relevant constraints.
In this version, {\rpathsim} takes a simple pattern as an input.
Hence, we supply the same pattern to both {\rpathsim} and PathSim, 
and compare their query processing times.
We use the same relationship patterns over DBLP and {\biomed} as described in
Section~\ref{sec:exp-robust}.
The average query processing times per query of {\rpathsim} and PathSim
are reported in Table~\ref{table:simple-efficiency} (under ``using Algorithm~\ref{alg:pathmod}'').
\ignore{
The average query processing time per query of {\rpathsim} (PathSim) 
over DBLP and {\biomed} dataset are 
0.034 (0.024) and 0.511 (0.477) seconds, respectively.
}
Overall, the running time of {\rpathsim} is slightly slower  
than PathSim due to the procedure of Algorithm~\ref{alg:pathmod}.
This result also shows that making {\rpathsim} more usable does {\it not} increase its running time considerably.

\begin{figure}[t]
\centering
\includegraphics[width=0.45\textwidth]{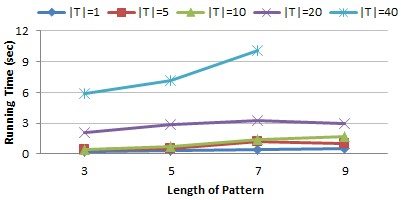}
\caption{Running times of {\rpathsim} in various settings}
\vspace{-10pt}
\label{fig:scale-runtime}
\end{figure}

We also evaluate the scalability of the version of {\rpathsim} that takes simple patterns as input as described in Section 5 over {\biomed} dataset. 
Since the running time of this version depends on 
the number of tgd constraints on the original database and the length of the input simple pattern, 
we measure its running time over different values of these parameters.
Because our datasets contain only two constraints, 
we test the scalability of our algorithm 
over a randomly generated set of constraints
from only {\biomed}. Af a matter of fact, we pick {\biomed} as it allows us to generate and test many constraints due to its rich structure.
We generate each constraint by using a coin flipping random generator
over each edge label in {\biomed} 
to decide whether an atom of the constraint contains such edge.
We limit the number of atoms in the premise of each constraint to be between 2 and 5,
and a single atom in its conclusion.
For the input pattern, 
each simple pattern is randomly generated such that 
it traverses the data graph from a drug to a disease.
We measure the running time by setting the 
number of constraints to 1, 5, 10, 20 and 40, 
and vary the length, i.e., number of edge labels, of the input simple pattern between 4 and 10.
We use the same query workload as the one described in Section~\ref{sec:exp-robust}. 
We report the average similarity query running times of {\rpathsim} 
per query over 5 runs of each setting in Figure~\ref{fig:scale-runtime}.
The Running time for pattern size of 9 for 40 constraints are omitted due to long running time.
Overall, {\rpathsim} is reasonably fast for a small number of constraints and up to the length of 8 for the input pattern.
The algorithm slows down when the number of constraint or the length of the input pattern goes up, 
especially when there are 20 or more constraints.
We also tested the version of RelSim in Section 5 without our proposed optimization techniques for ignoring non-relevant constraints.
The algorithm takes days to finish for 5 constraints or longer for more constraints for patterns of size more than 7.

\section{Related Work}
\label{sec:related}
There are robust algorithms over certain types of schematic variations 
\cite{Picado:2017:SIR:3035918.3035923,Tang:KDD:2017,Truong2012,ChodpathumwanAT16}. 
They, however, have two major shortcomings. First, they are robust only over a subset of 
frequently occurring schematic variations. 
Because they leverage the properties special to the variation 
over which they are robust, it is {\it not} clear how to generalize these algorithms 
to be robust against other schematic variations. 
Second, current schematically robust systems generally 
either propose new algorithms~\cite{Truong2012}, 
or make significant and/or complex modifications to the current ones  
\cite{Picado:2017:SIR:3035918.3035923}.     
However, current algorithms have 
been widely adapted and it is costly to replace them with new ones.
Hence, one should aim at making current algorithms robust to schematic variations using simple modifications.
Moreover, current algorithms are shown to be effective over some data representations. 
Thus, a robust version of them will be effective over more representations.
\bibliographystyle{ACM-Reference-Format}
\bibliography{ref}

\end{document}